\documentclass[aps,10pt,prd,notitlepage,tightenlines,nofootinbib,twocolumn,superscriptaddress]{revtex4-2}

\usepackage{amsmath}
\usepackage{amssymb,amsthm}
\usepackage{mathtools}
\usepackage{mathrsfs}
\usepackage{relsize}
\usepackage{bm}

\usepackage{epsfig,graphics,graphicx,color,xcolor}
\usepackage{soul} %strike out some texts  \st{something}
\usepackage{cancel} %cross out some texts \cancel{sth} \xcancel{sth}
\usepackage{slashed}	%feynman slashed symbols
\usepackage{subfigure}
\usepackage{array}
\usepackage{multirow}

\usepackage{ragged2e}
\usepackage{lineno}
\usepackage{natbib}
\usepackage{hyperref}
\hypersetup{colorlinks=true,linkcolor=red,anchorcolor=red,citecolor=orange, filecolor=brown,urlcolor=red,bookmarksnumbered=true,
pdfview=FitB
}
\graphicspath{ {image/} }
\usepackage{enumitem}

\colorlet{darkgreen}{green!50!black}
\colorlet{brightyellow}{yellow!75!red}
\colorlet{orange}{red!50!yellow}
\colorlet{darkgray}{gray!50!black}

\def\dd{{\mathrm{d}}}

\makeatletter
\newcommand*{\transpose}{%
  {\mathpalette\@transpose{}}%
}
\newcommand*{\@transpose}[2]{%
  \raisebox{\depth}{$\m@th#1\intercal$}%
}
\makeatother 
\newcommand{\cev}[1]{\reflectbox{\ensuremath{\vec{\reflectbox{\ensuremath{#1}}}}}}



\begin{document}

%opening 
\title{Spin-orbit correlation of quarks within quarkonium}

\author{Tianyang Hu}
\affiliation{Institute of Modern Physics, Chinese Academy of Sciences, Lanzhou 730000, China}
\affiliation{School of Nuclear Science and Technology, University of Chinese Academy of Sciences, Beijing 100049, China}

\author{Xianghui Cao}
\affiliation{Department of Modern Physics, University of Science and Technology of China, Hefei, Anhui 230026, China}

\author{Siqi Xu}
\affiliation{Institute of Modern Physics, Chinese Academy of Sciences, Lanzhou 730000, China}
\affiliation{School of Nuclear Science and Technology, University of Chinese Academy of Sciences, Beijing 100049, China}
\affiliation{Department of Physics and Astronomy, Iowa State University, Ames, Iowa 50010, U.S.}

\author{Weijie Du}
\affiliation{Institute of Modern Physics, Chinese Academy of Sciences, Lanzhou 730000, China}
\affiliation{School of Nuclear Science and Technology, University of Chinese Academy of Sciences, Beijing 100049, China}

\author{Qin-Tao Song}
\affiliation{School of Physics, Zhengzhou University, Zhengzhou, Henan 450001, China}

\author{Yang Li}
\thanks{Corresponding author}
\affiliation{Department of Modern Physics, University of Science and Technology of China, Hefei, Anhui 230026, China}
\affiliation{Anhui Center for Fundamental Sciences in Theoretical Physics, Hefei, 230026, China}

\date{\today}

\begin{abstract}
The spin-orbit correlation (SOC) provides a unique probe into the internal spin structure of hadrons. Defined via the parity-odd (P-odd) energy-momentum tensor (EMT), this observable can remain non-vanishing even in systems where the total angular momentum is zero. In this study, we connect the formal field-theoretical definition of the SOC to a non-perturbative quantum many-body framework utilizing light-front dynamics. Furthermore, we show how the SOC can be extracted directly from the hadronic matrix elements of the P-odd EMT, establishing a pathway to access this observable within the partonic picture. As a practical application, we compute the transverse and longitudinal SOC distributions for charmonium and $B_c$ mesons. While our findings align with rough estimates based on the Clebsch-Gordan decomposition, we demonstrate that these observables yield rich, non-trivial information regarding partonic dynamics.
\end{abstract}

 \maketitle

\section{Introduction} \label{sec:introduction}

Understanding the distribution of spin and orbital angular momentum (OAM) within hadrons stands as a central challenge in quantum chromodynamics (QCD) \cite{Gross:2022hyw}. Driven largely by the longstanding proton spin puzzle \cite{Ashman:1987hv}, decades of experimental campaigns have established that quark and gluon OAM are essential, yet elusive, components of the hadronic spin sum rule \cite{Adams:1994zd, SpinMuonSMC:1997voo, Alekseev:2010ub, Adare:2014hsq, Aidala:2012mv, Leader:2013jra, Accardi:2012qut}. As the field advances toward precision tomographic mapping at the future Electron-Ion Collider (EIC) via generalized parton distributions (GPDs) and transverse-momentum-dependent distributions (TMDs) \cite{AbdulKhalek:2021gbh, Aschenauer:2017jsk}, it is equally crucial to develop clear theoretical observables that isolate specific spin-orbit dynamics. Extending these theoretical investigations beyond the proton to systems like heavy quarkonia provides a pristine environment to investigate the underlying quantum many-body dynamics and fundamental spin-orbit correlations.

A critical step toward understanding hadronic spin lies in the energy-momentum tensor (EMT) $T^{\mu\nu}$, which rigorously encodes the spatial distribution of energy, momentum, and stress within the hadron \cite{Polyakov:2018zvc, Burkert:2018bqq}. While the standard parity-even EMT is foundational for extracting total angular momentum via Ji's sum rule \cite{Ji:1996ek}, decomposing this total $J^{q,g}$ into distinct spin and OAM contributions remains theoretically subtle due to scheme and gauge dependencies \cite{Wakamatsu:2010qj, Hatta:2011ku}. To circumvent these ambiguities and directly probe the interplay between spin and spatial dynamics, one can investigate the parity-odd (P-odd) EMT. Unlike its P-even counterpart, the P-odd EMT defines a robust spin-orbit correlation (SOC) observable that cleanly isolates the quantum entanglement between spin and orbit \cite{Thomas:1926dy}. Evaluating this correlation naturally bridges formal field-theoretical operators with non-perturbative tools, such as light-front wave function decompositions \cite{Brodsky:2000ii} and Wigner distributions \cite{Lorce:2011kd}.

At a fundamental level, SOCs $\vec L \cdot \vec S$ are ubiquitous across physical systems, from atomic fine structure to nuclear shell models, and play a similarly critical role in QCD. Remarkably, even in spin-0 hadrons such as the pion or $B_c$, a nontrivial spin-orbit structure can arise through the correlation $\langle L_z S_z\rangle$. Despite the vanishing total angular momentum of these states, this correlation encodes rich internal partonic dynamics \cite{Lorce:2011kd, Lorce:2013pza, Kanazawa:2014nha, Rajan:2017cpx, Engelhardt:2021kdo, Bhattacharya:2024sno, Bhattacharya:2024wtg, Benic:2025xpg, Agrawal:2025yoe}. This concept was first articulated by Lorcé for the nucleon \cite{Lorce:2014mxa, Bhoonah:2017olu} and subsequently extended to spin-0 systems \cite{Lorce:2025ayr}, prompting explicit calculations for both mesons \cite{Tan:2021osk, Acharyya:2024enp, Choudhary:2026zbe} and the proton \cite{Engelhardt:2021kdo}. Formally, this SOC is intimately tied to the aforementioned P-odd EMT, defined as $\widetilde{T}^{\mu\nu} = T_R^{\mu\nu} - T_L^{\mu\nu}$, where $T_{L,R}^{\mu\nu}$ represent the EMTs associated with left- and right-handed quark fields, respectively. Arising from the chiral separation of the quark EMT, the matrix elements of $\widetilde{T}^{\mu\nu}$ between hadronic states directly map the spatial distribution of the spin-orbit coupling, providing a direct link to the anomalous tensor charge and chiral-odd gravitational form factors (FF) \cite{Lorce:2014mxa}.

At the parton level, the SOC is encoded within twist-2 Wandzura-Wilczek contributions alongside genuine twist-3 structures \cite{Hatta:2024otc}. These correlations are central to novel momentum sum rules, analogous to the Jaffe-Manohar spin sum rule, that rigorously decompose hadronic energy into kinetic contributions and potential energies driven by color Lorentz forces \cite{Hatta:2024otc}. Furthermore, in the small-$x$ limit, the SOC exhibits a universal anti-alignment behavior: $C_q(x) \approx -\frac{1}{2}q(x)$ and $C_g(x) \approx -g(x)$. This striking asymptotic relation highlights a state of maximal quantum entanglement between the partonic spin and spatial degrees of freedom within this regime \cite{Bhattacharya:2024sno, Bhattacharya:2024wtg, Hatta:2024lbw}.

In this paper, we investigate spin-orbit coupling in heavy quarkonia, particularly charmonium ($c\bar c$) and $B_c$ ($b\bar c$) states, as ideal theoretical laboratories for probing hadronic spin structure \cite{Gross:2022hyw}. Quarkonia offer a controlled setting: at leading order, the sea quarks and gluons are suppressed, enabling clean separation of relativistic and nonrelativistic dynamics. Moreover, their mass scale permits complementary treatments via lattice QCD \cite{Dudek:2011tt}, effective field theories such as potential nonrelativistic QCD (pNRQCD) \cite{Brambilla:2004jw}, and light-front Hamiltonian approaches \cite{Li:2015zda, Li:2017mlw, Tang:2018myz, Tang:2019gvn, Wu:2026gul}.

This work is driven by two primary objectives. First, we derive a quantum many-body expression for the quark SOC density, $\mathcal C_z^q(r_\perp)$, within the framework of light-front dynamics. We further decompose the SOC along the longitudinal direction to establish a rigorous connection to its partonic interpretation, $C_z^q(x)$. Second, we explicitly compute $\mathcal C_z^q(r_\perp)$ for charmonia and $B_c$ states utilizing light-front wave functions obtained via basis light-front quantization (BLFQ, see Ref.~\cite{Vary:2025yqo} for a recent review). These evaluations yield quantitative predictions for future experiments and provide critical theoretical insights to help resolve the ongoing debate surrounding spin-orbit correlations in spin-0 hadrons. 
The remainder of the article is organized as follows. Section~\ref{sec:formalism} connects the field-theoretical SOC operator $\hat C_z^q$, defined via the P-odd EMT, to the canonical quantum many-body correlation $\sum_i S_i^z L_i^z$. In Sec.~\ref{sec:CLFD}, we analyze the covariant structure of the P-odd EMT hadronic matrix elements, demonstrating the proper extraction of both the SOC and the relevant physical FFs. Section~\ref{sec:partonic_interpretation} establishes the partonic interpretation of the SOC. We then present our explicit numerical calculations for heavy quarkonia in Sec.~\ref{sec:numerical_results}, before summarizing our conclusions in Sec.~\ref{sec:conclusion}.

\section{Spin-orbit correlation} \label{sec:formalism}

Within the framework of quantum many-body theory, the SOC is defined as
\begin{equation}C_{LS} = \langle \Psi | \sum_i \vec L_i \cdot \vec S_i | \Psi \rangle,
\end{equation}
where $\vec S$ and $\vec L$ denote the intrinsic spin and OAM, respectively. The quantum many-body state $|\Psi \rangle$ can be explicitly represented using the wave function $\Psi(\vec r_1, \vec r_2, \cdots, \vec r_n)$. To illustrate, in the hydrogen atom, $C_{LS} \approx 0$ for the 1S ground state, whereas $C_{LS} \approx -1$ for the P$_{1/2}$ state. The local spatial density of the SOC is formally obtained by inserting a Dirac delta function:
\begin{equation}
\mathcal C_{LS}(\vec r) = \langle \Psi | \sum_i \delta^3(\vec r - \vec r_i) \vec L_i \cdot \vec S_i | \Psi \rangle.
\end{equation}

Hadrons are inherently relativistic QCD bound states. To extend the concept of the SOC to QCD, we employ the light-front Hamiltonian formalism, where the complete quantum information of the system is encoded within the light-front wave functions (LFWFs). Because the total spin-orbit operator $\vec L\cdot \vec S$ is not Lorentz invariant, it is not a conserved quantity. However, in the infinite momentum frame, the transverse components of $\vec L$ and $\vec S$ are suppressed, leaving the longitudinal components invariant under longitudinal boosts. Consequently, it is sufficient to evaluate the SOC exclusively along the $z$-direction on the light front:
\begin{align}\label{eqn:LF_spin_orbit}
C_{LS}^z =\,& \langle \Psi|\sum_i L_{i}^z S_{i}^z |\Psi\rangle \\
=\, & \sum_n \sum_{\{s_i\}}\int[\dd x_i\dd^2 r_{i\perp}]_n 
  \widetilde\psi^*_n(\{x_i,\vec{r}_{i\perp}, s_i\}) \nonumber \\
   \times& \sum_j \eta_j s_j L^z_j 
  \widetilde\psi_n(\{x_i,\vec{r}_{i\perp}, s_i\}), \label{eqn:LFWF_spin_orbit}
\end{align}
Here, $\eta_j$ represents the intrinsic parity of the $j$-th parton, and $\eta_j = \pm 1$ for quarks and antiquarks, respectively; $s_j$ is the spin projection, and $L_{j}^z = -\hat z \cdot (\vec r_{j\perp} \times i\tensor \nabla_{j\perp})/2$ is the longitudinal OAM operator with $\tensor \nabla  =  \vec\nabla  - \cev \nabla$. 
The $n$-body LFWF, denoted by $ \widetilde\psi_n(\{x_i,\vec{r}_{i\perp}, s_i\})$, depends on both the transverse coordinates $\vec r_{i\perp}$ and the longitudinal momentum fractions $x_i = p^+_i/P^+$, where $p_i^+$ and $P^+$ are the longitudinal momenta of the $i$-th parton and the hadron, respectively. 
The light-front coordinates are defined as $x^\pm = x^0 \pm x^3$ and $\vec x_\perp = (x^1, x^2)$. The integration measure $[\dd x_i\dd^2 r_{i\perp}]_n$ spans the appropriate light-front phase space, as detailed in Appendix~\ref{sec:light-front_kinematics}. Thus, Eq.~(\ref{eqn:LFWF_spin_orbit}) establishes a direct LFWF representation of the SOC. 

Much like in the non-relativistic limit, transverse boosts are Galilean and strictly kinematical on the light front. This crucial property allows us to precisely localize the SOC in the transverse plane:
\begin{align}\label{eqn:LF_spin_orbit_density}
\mathcal C_{LS}^z(r_\perp) =\,& \langle \Psi|\sum_i \delta^2(r_\perp - r_{i\perp})L_{i}^z S_{i}^z |\Psi\rangle \\
=\, & \sum_n \sum_{\{s_i\}}\int[\dd x_i\dd^2 r_{i\perp}]_n 
  \widetilde\psi^*_n(\{x_i,\vec{r}_{i\perp}, s_i\}) \nonumber \\
 \times &  \sum_j \delta^2(r_\perp - r_{j\perp}) \eta_j s_j L^z_j 
  \widetilde\psi_n(\{x_i,\vec{r}_{i\perp}, s_i\}). \label{eqn:LFWF_spin_orbit_density}
\end{align}
Integrating this density over the transverse coordinates recovers the total SOC, 
\begin{equation}
C_{LS}^z = \int \dd^2r_\perp \, \mathcal C_{LS}^z(r_\perp).
\end{equation}
It should be noted that this integrated formulation does not yet resolve the distribution of $C_{LS}^z$ along the longitudinal axis. 

Bridging this framework to field theory, the quark contribution to the SOC is intimately linked to the P-odd EMT, as articulated in Ref.~\cite{Lorce:2014mxa}:
\begin{equation}\label{eqn:quark_spin_orbit_correlation}
\hat C_z^q = \frac{1}{2}\int \dd x^- \dd^2x_\perp \Big\{ x^1 \widetilde T^{+2}_q(x) - x^2\widetilde T^{+1}_q(x) \Big\}\Big|_{x^+=0},
\end{equation}
where, the P-odd kinematic EMT $\widetilde T^{\mu\nu}_q(x)$ is defined as, 
\begin{align}
 \widetilde T^{\mu\nu}_q(x) =\,& \frac{1}{2} \overline q(x) \gamma^{\mu} \gamma_5 i \tensor{D}^{\nu} q(x), \\
 =\,& T^{\mu\nu}_R - T^{\mu\nu}_L.
 \end{align}
Here $\tensor D^\mu = \vec\partial^\mu - \cev{\partial^\mu} - 2ig_sA^\mu$ is the covariant derivative, and 
$T^{\mu\nu}_{R,L} =(1/4)\overline q(x) \gamma^{\mu} (1 \pm \gamma_5) i \tensor{D}^{\nu} q(x)$ 
are the respective chiral projections, i.e. $T^{\mu\nu} = T^{\mu\nu}_R + T^{\mu\nu}_L$. It readily follows that the hadronic expectation value $C_z^q \equiv \langle \Psi | \hat C_z^q |\Psi\rangle$ obeys the exact same LFWF representation derived for $2C_{LS}^z$ in Eq.~(\ref{eqn:LFWF_spin_orbit_density}) when summing the quark and antiquark contributions, viz.
\begin{equation}
\hat C_z^q = 2\sum_i L_i^z S_i^z.
\end{equation}

To evaluate the spatial mapping of this correlation, we define the quark SOC density analogously to the integral operator in Eq.~(\ref{eqn:quark_spin_orbit_correlation}):
\begin{equation}\label{eqn:SOC_transverse_density}
\mathcal C_z^q(r_\perp) = r^1_\perp \widetilde{\mathcal T}^{+2}_q(\vec r_\perp) -  r^2_\perp \widetilde{\mathcal T}^{+1}_q(\vec r_\perp),
\end{equation} 
where, the light-front transverse density of the P-odd EMT in the Drell-Yan frame ($p'^+ = p^+$) is given by
\begin{equation}
\widetilde{\mathcal T}^{\mu\nu}_q(\vec r_\perp) = \int \frac{\dd^2 \Delta_\perp}{(2\pi)^22p^+}
e^{-i\vec \Delta_\perp \cdot \vec r_\perp} \langle p' |\widetilde T^{\mu\nu}_q(0) | p\rangle.
\end{equation}
Physically, the Drell-Yan kinematic condition $\Delta^+ \equiv p'^+ - p^+ = 0$ is equivalent to integrating out the longitudinal coordinate $x^-$. The factor of $2p^+$ naturally emerges from the normalization of the hadronic states, $\langle p' | p\rangle = 2p^+(2\pi)^3\delta^3(p-p')$. By construction, $\mathcal C_z^q(r_\perp)$ satisfies the LFWF representation established for $2\mathcal C_{LS}^z(r_\perp)$. 

Ultimately, this transverse density is the 2D Fourier transform of the hadronic matrix element $C_z^q(\Delta_\perp^2)$ evaluated in the Drell-Yan frame:
\begin{equation}
\mathcal C_z^q(b_\perp) = \int \frac{\dd^2 \Delta_\perp}{(2\pi)^2} e^{-i\vec \Delta_\perp \cdot \vec b_\perp} C_z^q(\Delta_\perp^2).
\end{equation}
The underlying hadronic matrix element is extracted via
\begin{equation}\label{eqn:SOC_HME}
C_z^q(\Delta^2_\perp) %\,&= \frac{1}{2P^+} \langle p' |\hat C_z^q | p\rangle \\
 = i \frac{\partial}{\partial \Delta_\perp^1} \frac{\langle p' | \widetilde T^{+2}_q(0)|p\rangle}{2P^+} - \big(1 \leftrightarrow 2\big)
\end{equation}
with average momentum $P = (p'+p)/2$ and momentum transfer $\Delta = p'-p$. $\Delta_\perp^2 \equiv \vec \Delta_\perp^2 = -\Delta^2$ in the Drell-Yan frame. The total quark SOC is subsequently determined by the forward limit, $C_z^q = C_z^q(0)$.

%%%%%%%%%%%%%%%%%%%%%%%%%%%%%

\section{Covariant decomposition}\label{sec:CLFD}

Following Eq.~(\ref{eqn:quark_spin_orbit_correlation}), the SOC is directly governed by the hadronic matrix element of the P-odd EMT, $\langle p' | \widetilde T^{\mu\nu}_q |p \rangle$. Dictated by Lorentz invariance, this matrix element can be parameterized in terms of fundamental P-odd covariant structures. For spin-0 hadrons, such as the scalar quarkonia investigated in this work, the Lorentz decomposition takes the form \cite{Lorce:2025ayr},
\begin{equation}
\langle p' | \widetilde T^{\mu\nu}_q(0) |p \rangle = -i \varepsilon^{\mu\nu\Delta P} \widetilde F_q(t).
\end{equation}
Here, $P = (p'+p)/2$, $\Delta = p'-p$, $t = -\Delta^2$, $\varepsilon^{\mu\nu\Delta P} = \varepsilon^{\mu\nu\rho\sigma}\Delta_\rho P_\sigma$, and $\varepsilon^{\mu\nu\rho\sigma}$ is the Levi-Civita tensor, with the convention $\varepsilon^{0123} = - \varepsilon_{0123} = +1$, which corresponds to  $\varepsilon^{-+12} = +2$ in light front coordinates. Note that $\varepsilon^{\mu\nu\Delta P}$ is anti-symmetric. 

In their study of pionic SOC \cite{Tan:2021osk}, Tan and Lyu augmented the standard Lorentz tensor $i \varepsilon^{\mu\nu\Delta P}$ with an additional P-odd tensor structure, ${P^{[\mu}\varepsilon^{\nu]+\Delta P}}$, a form originally proposed by Lorcé for the nucleon \cite{Lorce:2014mxa}. Because this term explicitly depends on the light-front ``$+$" direction, it fundamentally violates manifest Lorentz covariance for spin-0 hadrons, as recently emphasized by Lorcé and Song \cite{Lorce:2025ayr}. Nevertheless, such symmetry-breaking terms frequently manifest in practical calculations employing LFWFs that rely on finite Hamiltonian truncations, including Fock-space truncations\footnote{As demonstrated in Ref.~\cite{Cao:2024rul}, a similar breaking of the full Poincaré symmetry is also a known artifact in light-cone perturbation theory.}. Under such truncations, the full Lorentz group is explicitly broken down to the kinematic subgroup of the light front. To rigorously analyze and isolate these truncation-induced structures, one can employ covariant light-front dynamics (CLFD) \cite{Carbonell:1998rj, Cao:2024rul}. The CLFD framework explicitly tracks this frame dependence by introducing the light-front null vector $\omega^\mu = (1, 0, 0, -1)$, which dictates the orientation of light-front time ($\omega_\mu \omega^\mu = 0$, $\omega_\mu a^\mu = a^+$). It is particularly convenient to construct a scaled null vector $N^\mu = M^2\omega^\mu/(\omega \cdot P)$, where $M^2 = p^2 = p'^2$ denotes the squared hadron mass. This vector carries the appropriate mass dimension and remains manifestly invariant under arbitrary longitudinal scaling, $\omega^\mu \to \lambda \omega^\mu$. 

Incorporating this explicit $\omega$-dependence, the generalized covariant decomposition takes the form:
\begin{align}
\langle p' | \widetilde T^{[\mu\nu]}_q(0) |p \rangle = \,&  -i \varepsilon^{\mu\nu \Delta P} \widetilde F_q(t) 
 + \varepsilon^{\mu\nu PN}  \widetilde A_1(t) \nonumber \\
 & + i\varepsilon^{\mu\nu\Delta N} \widetilde A_2(t) \label{eq:P-odd_Tmunu_asym} \\
\langle p' | \widetilde T^{\{\mu\nu\}}_q(0) |p \rangle =\,& 
-iP^{\{\mu}\varepsilon^{\nu\} P\Delta N} \widetilde S_1(t) 
 + \Delta^{\{\mu}\varepsilon^{\nu\}P\Delta N} \widetilde S_2(t) \nonumber \\
& + iN^{\{\mu}\varepsilon^{\nu\} P \Delta N} \widetilde S_3(t) \label{eq:P-odd_Tmunu_sym}
\end{align}
where, $a^{[\mu}b^{\nu]} = (a^\mu b^\nu - a^\nu b^\mu)/2$, and $a^{\{\mu}b^{\nu\}} = (a^\mu b^\nu + a^\nu b^\mu)/2$.  
Consequently, spurious FFs arise due to the explicit dependence on the light-front orientation $N^\mu$, which violates manifest Lorentz covariance. Specifically, the FFs $\widetilde A_{1,2}(t)$ within the antisymmetric part of $\widetilde T^{\mu\nu}_q$, alongside $\widetilde S_{1,2,3}(t)$ within its symmetric part, are entirely spurious. These artifacts must strictly vanish upon the restoration of full Lorentz symmetry. Furthermore, the alternative P-odd structure proposed in Ref.~\cite{Tan:2021osk} is mathematically related to the basis structures identified above via the Schouten identities,
\begin{equation}
P^{[\mu}\varepsilon^{\nu] N \Delta P} = M^2 \varepsilon^{\mu\nu\Delta P} - P^2 \varepsilon^{\mu\nu\Delta N}. 
\end{equation}

The physical FF $\widetilde F_q(t)$ can be extracted from the antisymmetric part of the transverse components $\widetilde T^{+i}$. Evaluated in the symmetric Drell-Yan frame ($\Delta^+=0, \vec P_\perp = 0$), this expression reads
\begin{equation}\label{eqn:HME_Ftilde}
\langle p' | \widetilde T^{[+i]}_q(0) | p\rangle = -i P^+ \epsilon^{ij}_T \Delta^j_\perp \widetilde F_q(\Delta^2_\perp).
\end{equation}
Here $\epsilon^{ij}_T$ is the transverse antisymmetric tensor with $\epsilon^{12}_T = -\epsilon^{21}_T = +1$. 
Conversely, the symmetric part of $\widetilde T^{+i}_q$ may not identically vanish for model wave functions, and is directly related to the spurious FF $\widetilde S_1$:
\begin{equation}\label{eqn:HME_Stilde}
\langle p' | \widetilde T^{\{+i\}}_q(0) | p\rangle = -i M^2 P^+ \epsilon^{ij}_T \Delta^j_\perp \widetilde S_1(\Delta^2_\perp).
\end{equation}
The complete set of hadronic matrix elements is collected in Appendix~\ref{sec:hadron_matrix_elements}. Returning to the SOC, the FF $C_z^q(\Delta^2_\perp)$ is defined directly from the hadronic matrix element of the full transverse component $\widetilde T^{+i}_q = \widetilde T^{[+i]}_q + \widetilde T^{\{+i\}}_q$, which consequently incorporates the spurious contribution:
\begin{multline}
C_z^q(\Delta^2_\perp) = \widetilde F_q(\Delta_\perp^2) + \Delta^2_\perp \frac{\dd}{\dd \Delta^2_\perp}\widetilde F_q(\Delta^2_\perp) \\ + M^2 \widetilde S_1(\Delta_\perp^2) + M^2 \Delta^2_\perp \frac{\dd}{\dd \Delta^2_\perp}\widetilde S_1(\Delta^2_\perp). 
\end{multline}

Lorcé and Song showed that the FF $\widetilde F_q(t)$ is related to two other FFs \cite{Lorce:2025ayr},
\begin{equation}\label{eqn:FF5_and_F_H}
\widetilde F_q(t) = -\frac{1}{2} F^q(t) + \frac{m_q}{2M} H^q(t),
\end{equation}
a relation that follows directly from the QCD operator identity \cite{Lorce:2014mxa}
\begin{equation}
\bar q \gamma^{[\mu}\gamma_5 i \tensor D^{\nu]} q = m_q \bar q i \sigma^{\mu\nu}\gamma_5 q + \frac{1}{2}\varepsilon^{\mu\nu\alpha\beta}\partial_\alpha\big(\bar q \gamma_\beta q\big).
\end{equation}
Here, the FFs $F^q(t)$ and $H^q(t)$ are defined through the local hadronic matrix elements,
\begin{align}
& \langle p'|\bar q(0) \gamma^+ q(0)|p\rangle = 2P^+ F^q(t), \\
& \langle p'|\bar q(0) i\sigma^{+i}\gamma_5 q(0)|p\rangle = - \frac{i\varepsilon^{+i\Delta P}}{M}H^q(t).
\end{align}
Because $\widetilde F_q$ is extracted solely from the antisymmetric part of the hadronic matrix element according to Eq.~(\ref{eqn:HME_Ftilde}), this underlying relationship is naturally preserved even in the presence of spurious contributions.

Having established a rigorous procedure to isolate the physical FF $\widetilde F_q$ from the macroscopic EMT using CLFD, we now turn to the underlying microscopic picture. In the next section, we investigate the partonic interpretation of the SOC, bridging this formal matrix element to the internal dynamics of the bound state via Wigner distributions and GPDs.

\section{Partonic interpretation}\label{sec:partonic_interpretation}

The SOC admits a natural partonic interpretation in terms of polarized Wigner distributions \cite{Lorce:2014mxa}, 
\begin{equation}\label{eqn:wigner}
C_z^q = \int \dd x \dd^2 k_\perp \dd^2 b_\perp (\vec b_\perp \times \vec k_\perp)_z \widetilde f_q(x, \vec k_\perp, \vec b_\perp)
\end{equation}
where, $\widetilde f_q$ represents the polarized quark Wigner distribution:
\begin{multline}
\widetilde f_q(x, \vec k_\perp, \vec b_\perp) = 
\frac{1}{2}\int \frac{\dd^2\Delta_\perp}{(2\pi)^2} e^{-i\vec\Delta_\perp\cdot \vec b_\perp} \int \frac{\dd z^-\dd^2 z_\perp}{2(2\pi)^3} \\
\times e^{\frac{i}{2}xP^+z^--i\vec k_\perp\cdot \vec z_\perp} 
\langle p' | \bar q(-\frac{z}{2}) \gamma^+\gamma_5 W_\pm q(\frac{z}{2}) |p\rangle\big|_{z^+=0}. 
\end{multline}
In this expression, $W_\pm$ denotes a staple-shaped Wilson line connecting the two spacetime points $\pm z/2$ via light-front infinity at $z^- = \pm \infty$.

Retaining the explicit dependence on the impact parameter $\vec b_\perp$ in Eq.~(\ref{eqn:wigner}) yields the SOC transverse density $\mathcal C_z^q(b_\perp)$ introduced in Eq.~(\ref{eqn:SOC_transverse_density}). Similarly, integrating over the transverse phase space while leaving the longitudinal momentum fraction $x$ unintegrated yields the parton distribution of the SOC in the longitudinal direction:
\begin{equation}
C_z^q(x) = \int \dd^2 k_\perp \dd^2 b_\perp (\vec b_\perp \times \vec k_\perp)_z \widetilde f_q(x, \vec k_\perp, \vec b_\perp).
\end{equation}
Retaining the transverse momentum $\vec k_\perp$ in this integration naturally defines the transverse momentum dependent (TMD) parton distribution:
\begin{equation}
C_z^q(x, k_\perp) = \int  \dd^2 b_\perp (\vec b_\perp \times \vec k_\perp)_z \widetilde f_q(x, \vec k_\perp, \vec b_\perp).
\end{equation}

While the above definition is intuitive, it requires evaluating a gauge link in the transverse direction. Alternatively, the purely longitudinal distribution $C_z^q(x)$ can be determined using the GPD formalism \cite{Lorce:2025ayr}. Following Eqs.~(\ref{eqn:HME_Ftilde}--\ref{eqn:FF5_and_F_H}), $C_z^q(x)$ is directly related to the leading-twist GPDs $\mathcal H^q(x, \xi, t)$ and $\mathcal H^q_1(x, \xi, t)$, as well as the twist-3 GPD $\mathcal G_2^q(x, \xi, t)$:
\begin{equation}\label{eqn:Cz(x)}
 C_z^q(x) =  \frac{2m_q}{M} \mathcal H_1^q(x, 0, 0)  - \mathcal H^q(x, 0, 0) + x \mathcal G_2^q(x, 0, 0),
 \end{equation}
 where these GPDs are defined in terms of bilocal light-cone operators:
 \begin{align}
  & \int \frac{\dd z^-}{4\pi} e^{\frac{i}{2} xP^+z^-} \langle p'|\bar q(-z/2)\gamma^+ q(z/2)|p\rangle \big|_{z^+=z_\perp=0} \nonumber \\
  & \quad = 2 \mathcal H^q(x, \xi, t), \\
 & \int \frac{\dd z^-}{4\pi} e^{\frac{i}{2} xP^+z^-} \langle p'|\bar q(-z/2)i \sigma^{i+}\gamma_5 q(z/2)|p\rangle \big|_{z^+=z_\perp=0} \nonumber\\
  & \quad= \frac{2i\epsilon^{ij}\Delta^j}{M} \mathcal H_1^q(x, \xi, t), \\
  & \int \frac{\dd z^-}{4\pi} e^{\frac{i}{2} xP^+z^-} \langle p'|\bar q(-z/2) \gamma^i\gamma_5 q(z/2)|p\rangle \big|_{z^+=z_\perp=0} \nonumber\\
  & \quad= -\frac{i\epsilon^{ij}\Delta^j}{P^+} \mathcal G_2^q(x, \xi, t).
 \end{align}
From these definitions, the LFWF representation of $C_z^q(x)$ can be readily obtained. Furthermore, if the spurious contributions vanish, the twist-3 GPD $\mathcal G_2^q$ is rigorously related to the leading-twist GPDs $\mathcal H^q$ and $\mathcal H_1^q$ via the QCD equations of motion. In this limit, $C_z^q(x)$ depends exclusively on $\mathcal G_2^q$, as first demonstrated in Ref.~\cite{Lorce:2025ayr}.

\section{Numerical results}\label{sec:numerical_results}

In this section, we evaluate the SOC for charmonia and $B_c$ states. Because the valence Fock sector dominates the wave function of heavy quarkonia \cite{Shi:2021nvg, Cao:2025bit}, we evaluate the observables within this leading sector. Truncating Eq.~(\ref{eqn:LFWF_spin_orbit}) accordingly, the valence contribution to the SOC takes the explicit form:
\begin{multline}
C_z^q(x) = \frac{1}{2\pi} \sum_{s\bar s}\int \dd^2 r_\perp \big|\widetilde \psi_{s\bar s}(x, \vec r_\perp)\big|^2 \\
\times \big\{ s(1-x) - \bar s x\big\} l^z.
\end{multline}
Notably, because the relevant operator $\hat C_z^q$ is odd under charge conjugation, the total combined SOC identically vanishes for charge-symmetric states like charmonia. Consequently, our analysis for these states focuses exclusively on the isolated quark (or antiquark) contributions. Conversely, for asymmetric systems such as the $B_c$ meson, the total SOC is generally non-vanishing. In the small-$x$ limit, the valence expression asymptotically reduces to:
\begin{equation}
C_z^q(x) \approx  C_z^q q(x).
\end{equation} 

We adopt heavy quarkonium wave functions obtained from the BLFQ approach \cite{Li:2017mlw, Tang:2018myz}. In this framework, wave functions are determined by diagonalizing an effective light-front Hamiltonian within a holographic basis. Practical calculations require a basis truncation, which is governed by the cutoff parameter $N_\text{max}$. This truncation dictates the effective ultraviolet (UV) resolution via $\Lambda_\text{UV}\sim\kappa\sqrt{N_\text{max}}$, where $\kappa\approx 1\ \text{GeV}$ is the basis scale parameter. Guided by our previous studies \cite{Li:2017mlw, Tang:2018myz, Tang:2020org, Li:2021ejv, Wang:2023nhb}, we select $N_\text{max}=8$ to compute the central values for charmonium, which corresponds to a UV resolution comparable to the charmonium mass ($\Lambda_\text{UV}\approx M$). The sensitivity to this basis truncation is then estimated by the difference between the $N_\text{max}=8$ and $N_\text{max}=16$ results. For the $B_c$ system, we similarly adopt the $N_\text{max}=32$ results for our central values, while the basis sensitivity is estimated as twice the difference between the $N_\text{max}=32$ and $N_\text{max}=24$ results. The specific values of the BLFQ model parameters utilized for each hadronic system are detailed in Refs.~\cite{Li:2017mlw, Tang:2018myz}.

Table~\ref{tab:SOC} summarizes the charm quark SOC for several low-lying charmonia, alongside the total SOC for the $B_c$ states. A naïve estimate based on the non-relativistic quark model (NRQM) is also presented as a baseline. As seen in the table, the SOC of the quark is consistently negative, implying that the spin and OAM of the quark are anti-aligned within the quarkonium system. Conversely, the SOC of the antiquark is positive, owing to the explicit negative sign associated with the antiquark in the definition. Consequently, charge conjugation symmetry causes the total SOC for charmonia to vanish exactly. 

Furthermore, Table~\ref{tab:SOC} reveals that the BLFQ results for S-wave states are not exactly zero, in stark contrast to expectations from the NRQM. This deviation is a distinct relativistic effect. Specifically, it stems from dynamically generated P-wave mixing, a feature forbidden by non-relativistic symmetry rules but naturally accommodated within relativistic light-front dynamics \cite{Li:2017mlw}. For P-wave and D-wave mesons, our BLFQ results are of the same order of magnitude as the NRQM estimates, which is consistent with the predominantly non-relativistic nature of these heavy bound states. 

\begin{table}
\caption{
Spin-orbit correlation (SOC) $C_z^q$ for selected charmonia and $B_c$ states, obtained using basis light-front quantization (BLFQ). Quoted uncertainties reflect the theoretical sensitivity to the basis truncation (see text). For charmonia, only the quark contribution is presented, as the total SOC vanishes exactly due to charge conjugation symmetry. For the $B_c$ system, the total SOC is shown. The final column provides theoretical estimates derived from Clebsch-Gordan coefficients within the non-relativistic quark model (NRQM).}
\label{tab:SOC}
\begin{tabular}{c|ccc}
\toprule
 &		& \textbf{BLFQ} 		 & \textbf{NRQM} \\
\hline 
\multirow{17}{*}{$m_J=0$} & $\eta_c(1S)$	& $-0.04(1)$	 & 	0 \\
& $\eta_c(2S)$	& $-0.019(16)$ &	0 \\
& $\eta_c(3S)$	& $-0.006(10)$ &	0 \\
& $\chi_{c0}(1P)$	& $-0.367(5)$	 &	$-1/3$ \\
& $\chi_{c0}(2P)$ & $-0.414(18)$ & 	$-1/3$ \\
& $J/\psi$ &	$-0.0003(0)$	 &	0 \\
& $\psi(2S)$   &	$-0.0010(1)$   &	0 \\
& $\psi(3S)$  &	$-0.0008(4)$   &	0 \\
& $\psi(1D)$  & $-0.3799(86)$ &	$-3/10$ \\
& $\chi_{c1}(1P)$  &	$-0.457(15)$	& $-1/2$ \\
& $\chi_{c1}(2P)$  &	$-0.468(28)$  	& $-1/2$ \\
& $h_c(1P)$  &	$-0.0125(18)$ 	&	0 \\
& $h_c(2P)$  &	$-0.0071(42)$ 	& 	0 \\ \cline{2-4}
& $B_c(1S)$ &	$0.041(6)$	&	0 \\
& $B_c(2S)$ &	$0.04(1)$ 	&	0 \\
& $B_c(1P)$ &	$0.343(6)$	&	$0.33$ \\
& $B_c(2P)$ &	$0.356(11)$  	&	$0.33$ \\
\hline 
\multirow{8}{*}{$m_J=1$}  &
$J/\psi$  	& $-0.0019(7)$ &	0 \\
& $\psi(2S)$   &	 $-0.0020(4)$ &	0 \\
& $\psi(1D)$   & $-0.8432(35)$ &	$-3/5$ \\
& $\psi(3S)$   & $-0.0006(6)$	&	0 \\
& $\chi_{c1}(1P)$  	& $-0.0174(17)$ &	0 \\
& $\chi_{c1}(2P)$  	& $-0.0105(33)$ & 0 \\
& $h_c(1P)$  	& $-0.0179(29)$ &	0 \\
& $h_c(2P)$  	& $-0.009(5)$   &	0 \\
\botrule
\end{tabular}
\end{table}

\begin{figure*}
    \centering
    \subfigure[\label{fig:SOC_etacchic0Bc_rCofaq_1S1P_N816_coordinates}]{\includegraphics[width=0.4\textwidth]{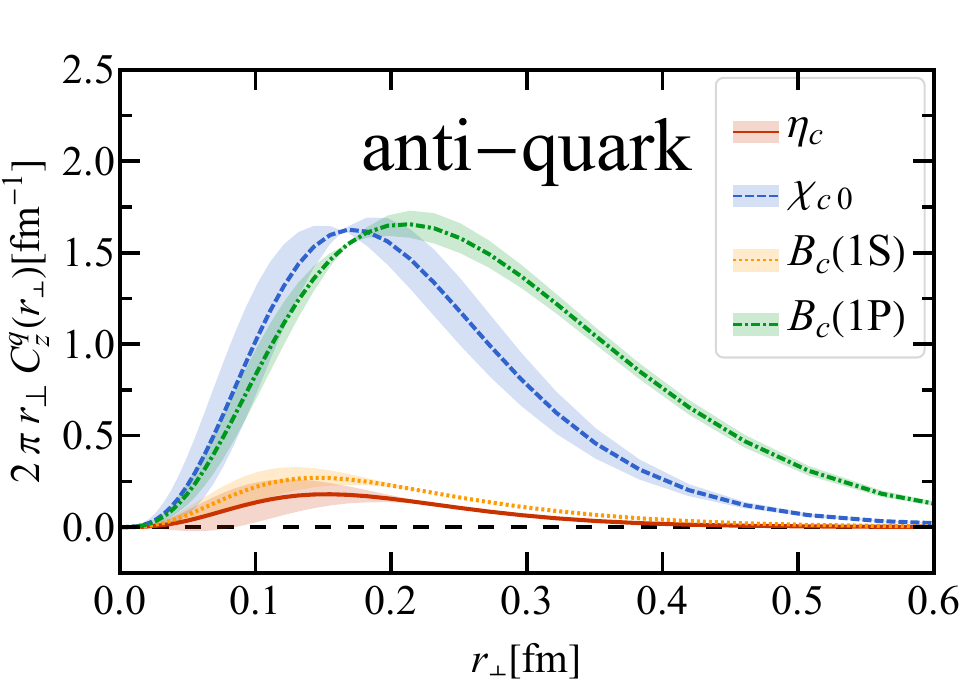}}
    \subfigure[\label{fig:SOC_etacchic0Bc_rCofq_1S1P_N816_coordinates}]{\includegraphics[width=0.41\textwidth]{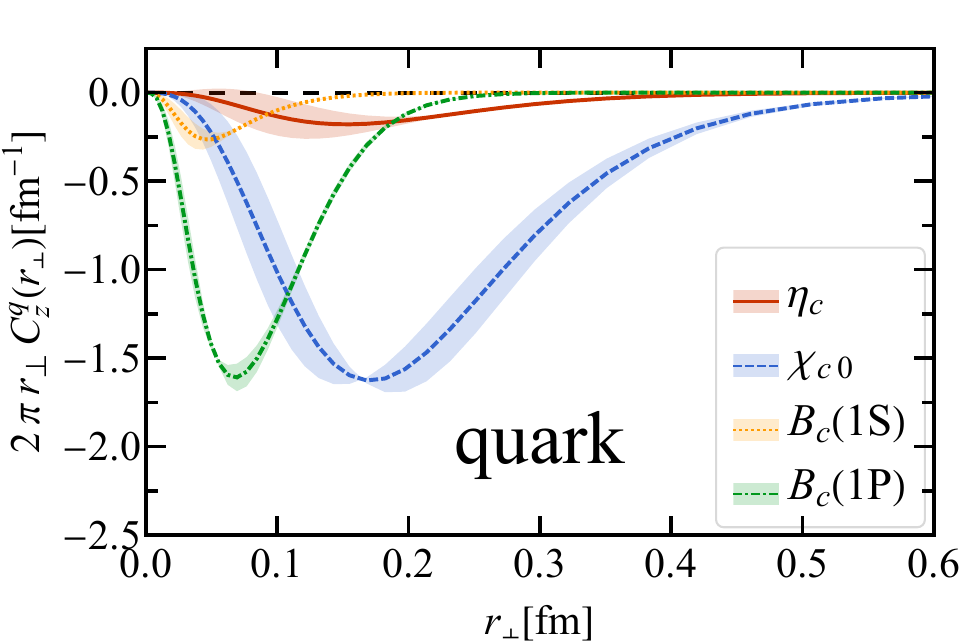}}
    \caption{
Spin-orbit correlation distributions for the ground and excited states $\eta_c$, $\chi_{c0}$, $B_c(1S)$, and $B_c(1P)$. Panel (a) illustrates the isolated antiquark contribution, whereas panel (b) displays the corresponding quark contribution. The accompanying shaded bands reflect the theoretical sensitivity to the basis truncation (see text for details). 
    }
    \label{fig:SOC_etacchic0Bc_rCofq_1S1P_N2432}
\end{figure*}

\begin{figure*}
    \centering
    \subfigure[\label{fig:SOC_etac_rCofc_1S2S3S_N816_coordinates}]{\includegraphics[width=0.405\textwidth]{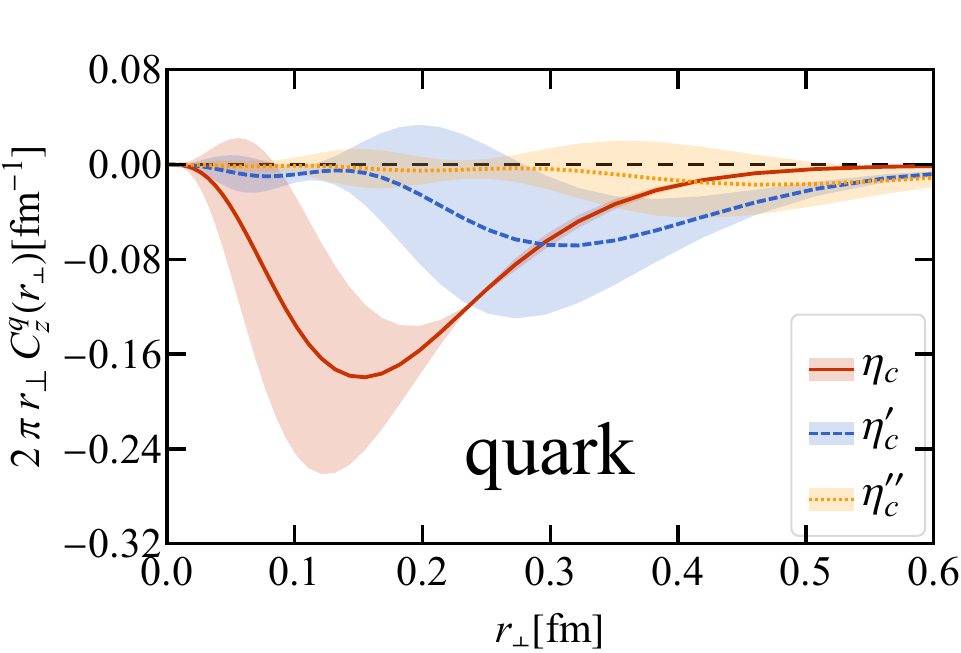}}
    \subfigure[\label{fig:SOC_chic0_rCofc_1P2P_N816_coordinates}]{\includegraphics[width=0.4\textwidth]{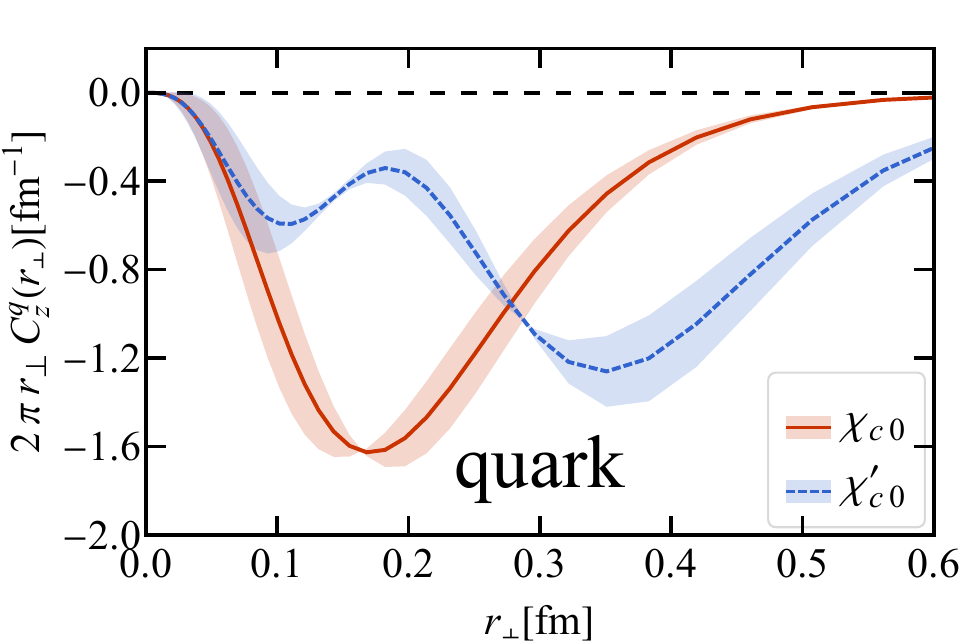}}
    \subfigure[\label{fig:SOC_Bc_rC_1S2S_N2432_coordinates}]{\includegraphics[width=0.405\textwidth]{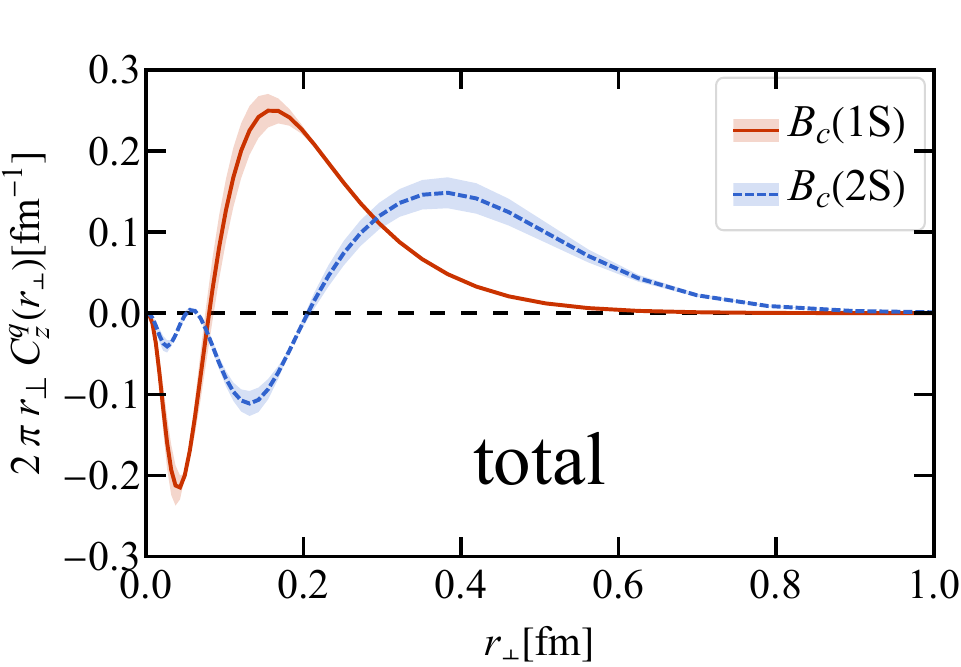}}
    \subfigure[\label{fig:SOC_Bc_rC_1P2P_N2432_coordinates}]{\includegraphics[width=0.4\textwidth]{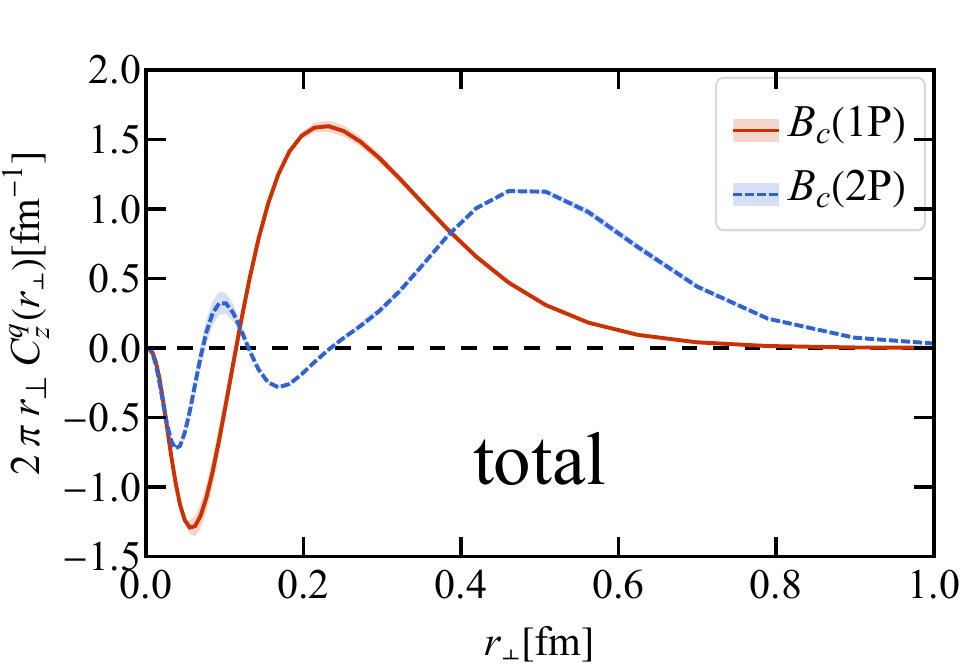}}
        \caption{
        Transverse distribution of the quark spin-orbit correlation, $2\pi r_\perp \mathcal C^q_z(r_\perp)$, for (a) S-wave charmonia ($\eta_c, \eta_c'$, and $\eta_c''$); (b) P-wave charmonia ($\chi_{c0}$ and $\chi_{c0}'$); (c) S-wave $B_c$ mesons ($B_c(1S), B_c(2S)$), and (d) P-wave scalar $B_c$ mesons ($B_c(1P), B_c(2P)$). The shaded bands represent the theoretical uncertainty due to basis sensitivity (see text for details).         }
    \label{fig:SOC_etacchic0Bc_rC_SP}
\end{figure*}

The spatial distribution of the SOC in the transverse plane, $\mathcal C_z^q(r_\perp)$, is visualized in Figs.~\ref{fig:SOC_etacchic0Bc_rCofq_1S1P_N2432}--\ref{fig:SOC_etacchic0Bc_rC_SP}. Figure~\ref{fig:SOC_etacchic0Bc_rCofq_1S1P_N2432} explicitly compares these transverse densities across a selection of ground and orbitally excited states, namely $\eta_c(1S)$, $\chi_{c0}(1P)$, $B_c(1S)$, and $B_c(1P)$. Consistent with physical intuition, the magnitude of the SOC is substantially enhanced in the P-wave systems relative to their S-wave counterparts. Extending this analysis to radial excitations, Fig.~\ref{fig:SOC_etacchic0Bc_rC_SP} illustrates the corresponding transverse densities for higher radially excited quarkonia. In these states, a clear nodal structure manifests within the density, serving as a direct physical imprint of the alternating signs in the radially excited wave functions.

\begin{figure*}
    \centering
    \subfigure[\label{fig:SOCPDF_etac_C_1S_N816}]{\includegraphics[width=0.4\textwidth]{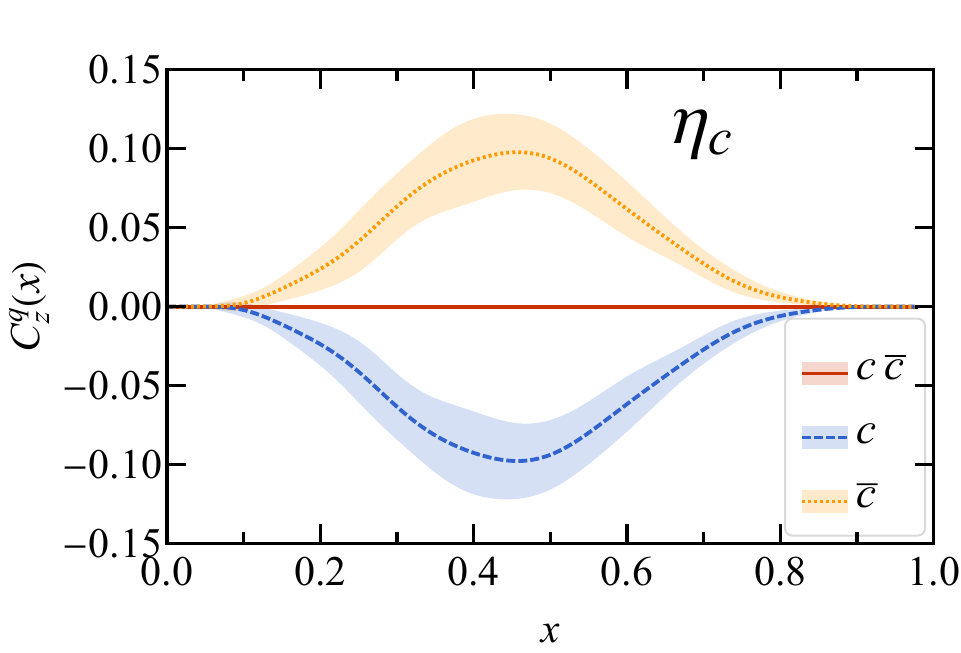}}
    \subfigure[\label{fig:SOCPDF_etac_C_2S_N816}]{\includegraphics[width=0.41\textwidth]{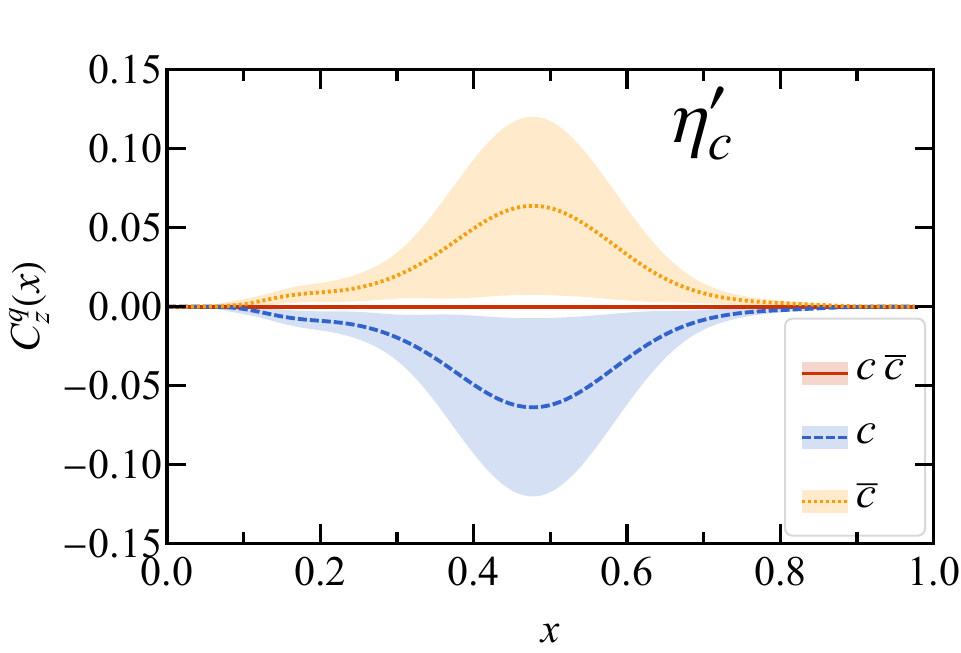}}
    \subfigure[\label{fig:SOCPDF_chic0_C_1P_N816}]{\includegraphics[width=0.41\textwidth]{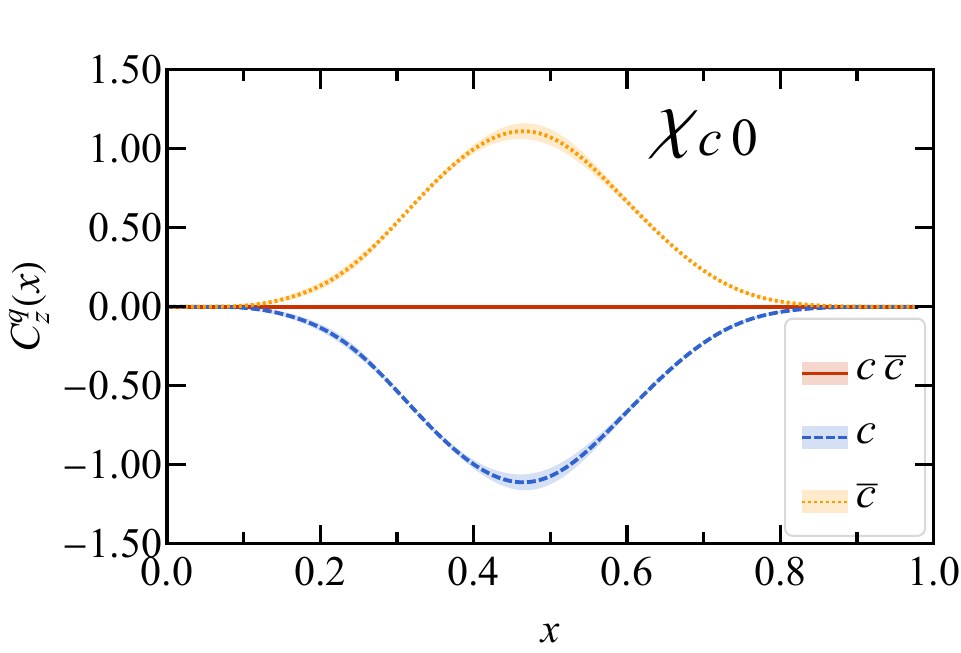}}
    \subfigure[\label{fig:SOCPDF_chic0_C_2P_N816}]{\includegraphics[width=0.41\textwidth]{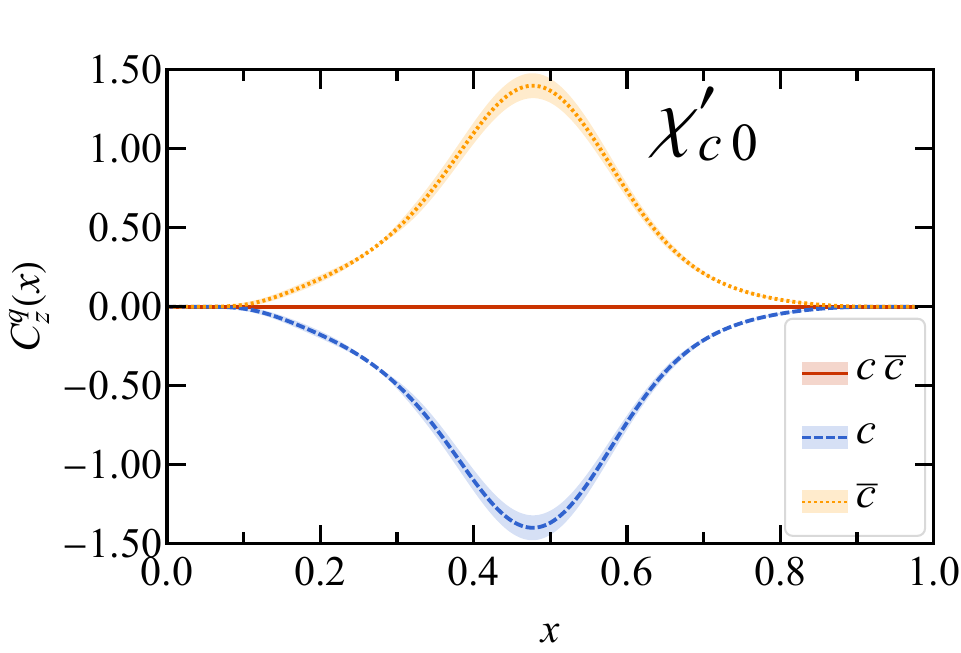}}
    \subfigure[\label{fig:SOCPDF_Bc_C_1S_N2432}]{\includegraphics[width=0.41\textwidth]{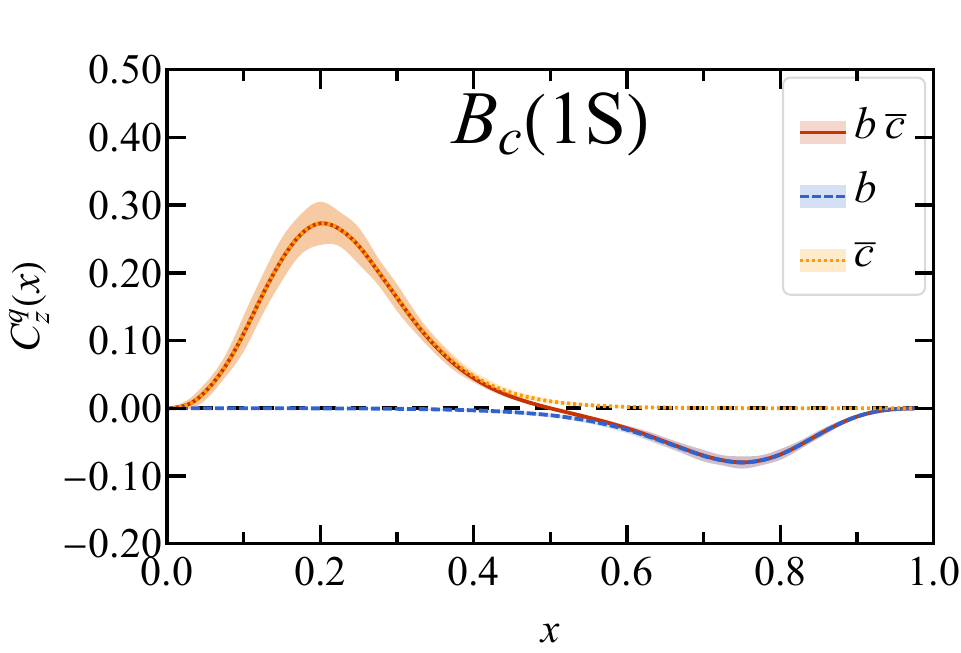}}
    \subfigure[\label{fig:SOCPDF_Bc_C_2S_N2432}]{\includegraphics[width=0.41\textwidth]{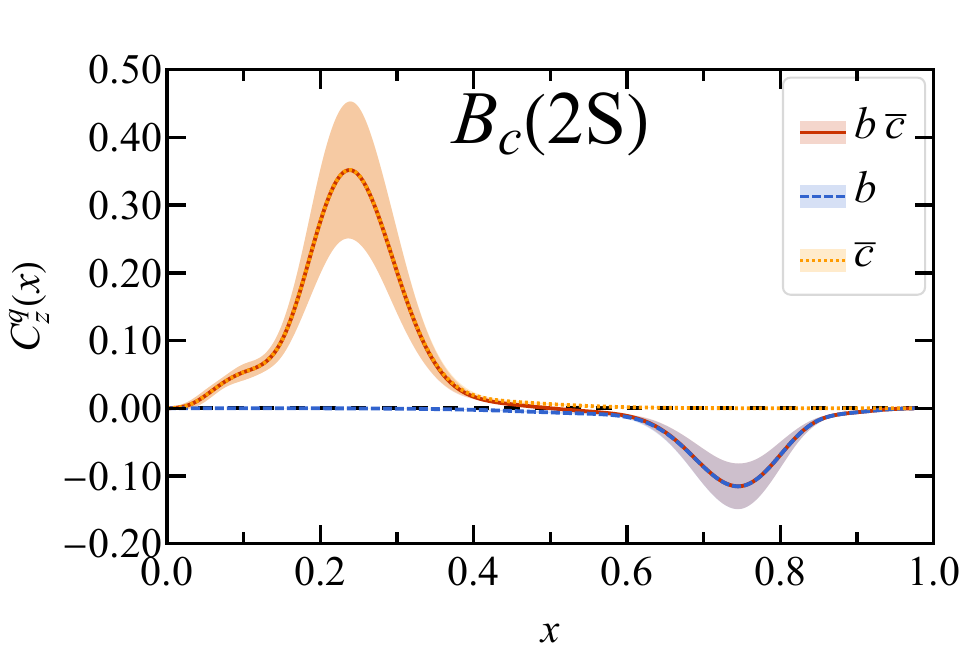}}
    \subfigure[\label{fig:SOCPDF_Bc_C_1P_N2432}]{\includegraphics[width=0.41\textwidth]{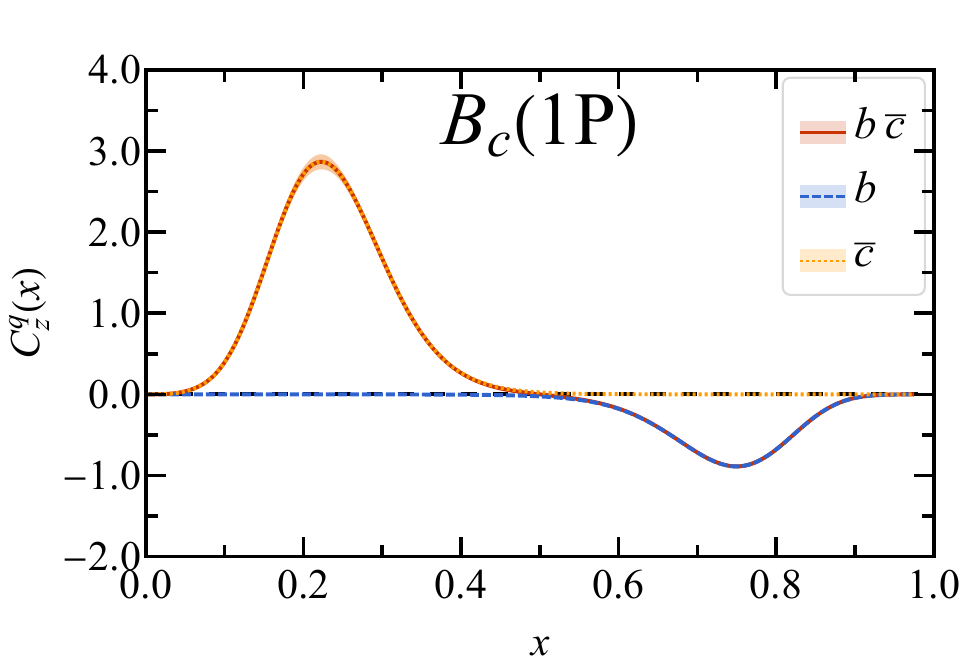}}
    \subfigure[\label{fig:SOCPDF_Bc_C_2P_N2432}]{\includegraphics[width=0.41\textwidth]{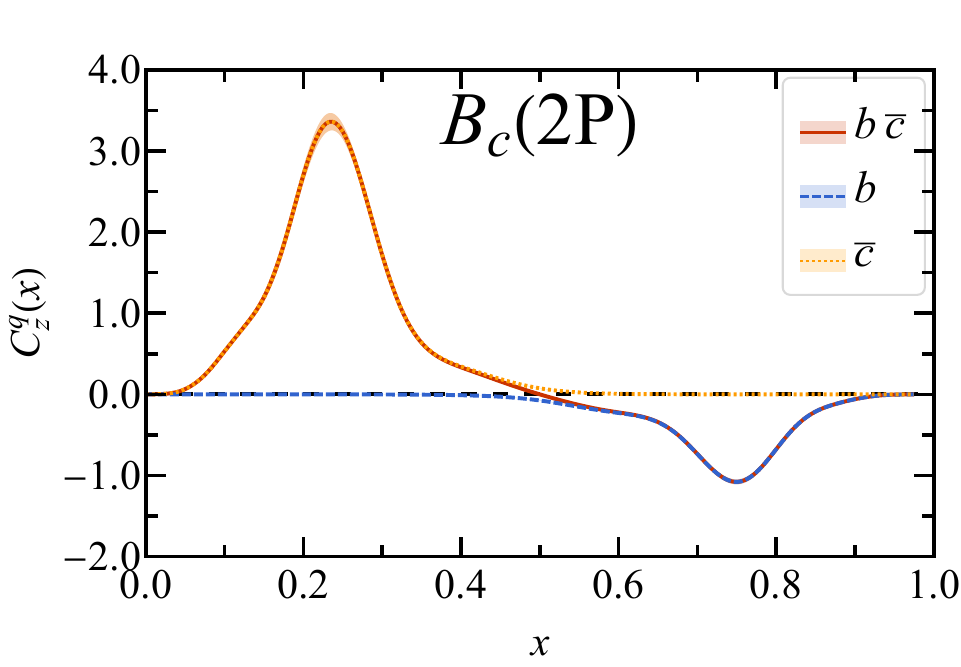}}
    \caption{
    Parton distribution of the spin-orbit correlation, $C^q_z(x)$, for (a) $\eta_c$; (b) $\eta'_c$; (c) $\chi_{c0}$; (d) $\chi_{c0}'$; (e) $B_c(1S)$; (f) $B_c(2S)$; (g) $B_c(1P)$; and (h) $B_c(2P)$. The shaded bands represent the theoretical uncertainty due to basis sensitivity (see text for details).}
    \label{fig:SOCPDF}
\end{figure*}

\begin{figure*}
    \centering
    \subfigure[\label{fig:SOCPDF_etac_Cofq_1S_N8_coordinates}]{\includegraphics[width=0.35\textwidth]{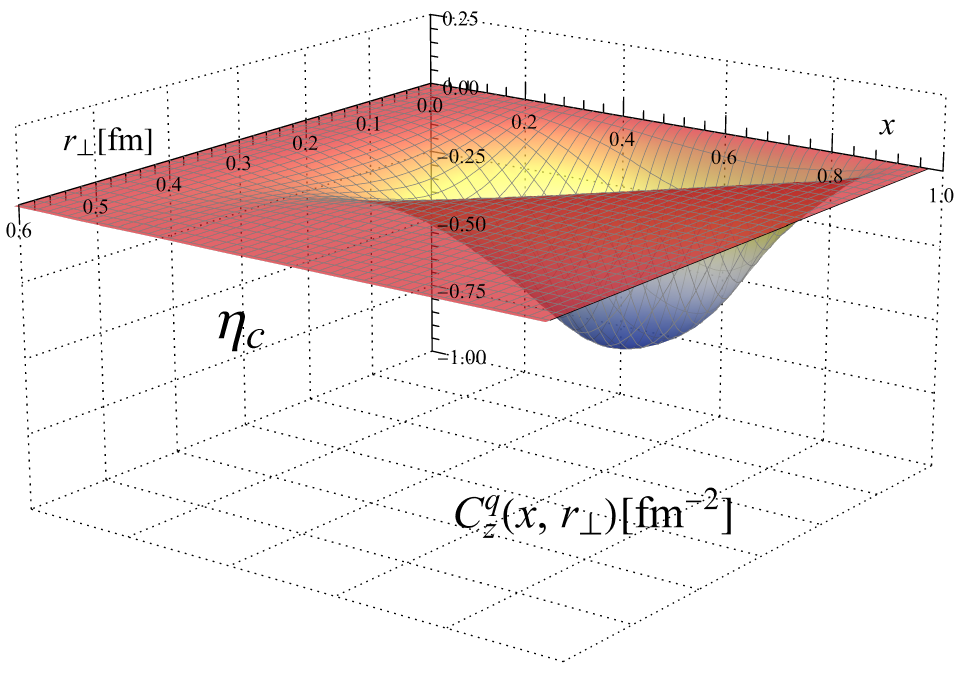}}
    \subfigure[\label{fig:SOCPDF_etac_Cofq_2S_N8_coordinates}]{\includegraphics[width=0.35\textwidth]{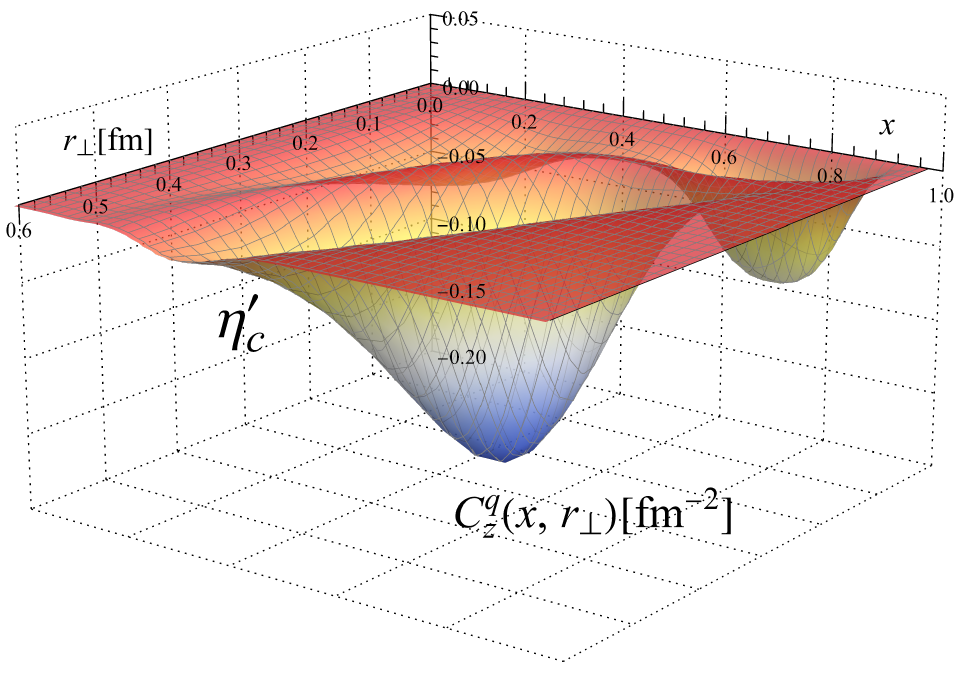}}
    \subfigure[\label{fig:SOCPDF_chic0_Cofq_1P_N8_coordinates}]{\includegraphics[width=0.35\textwidth]{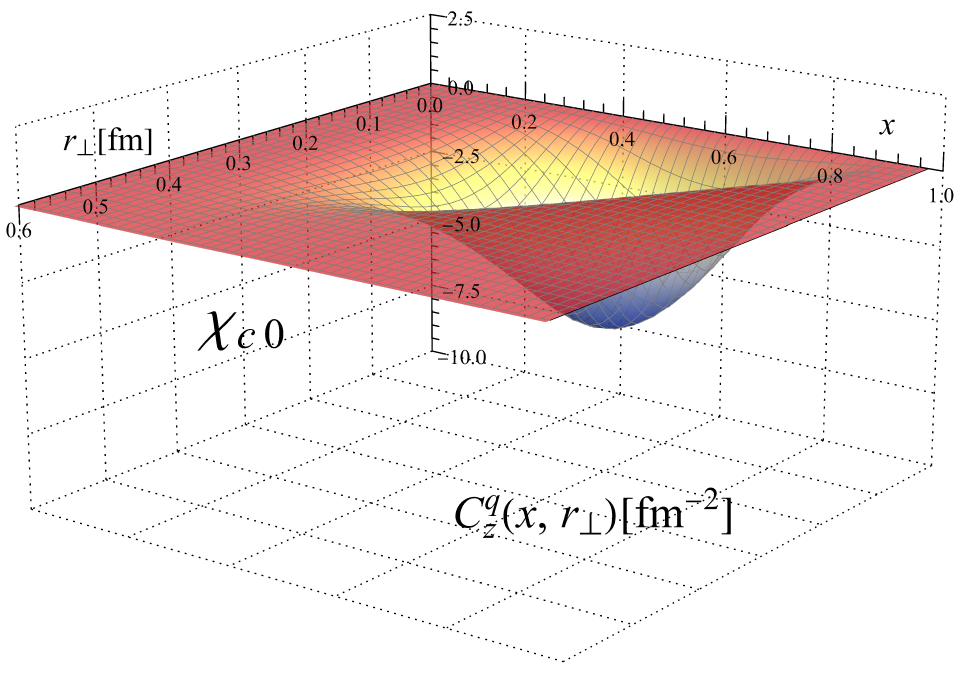}}
    \subfigure[\label{fig:SOCPDF_chic0_Cofq_2P_N8_coordinates}]{\includegraphics[width=0.35\textwidth]{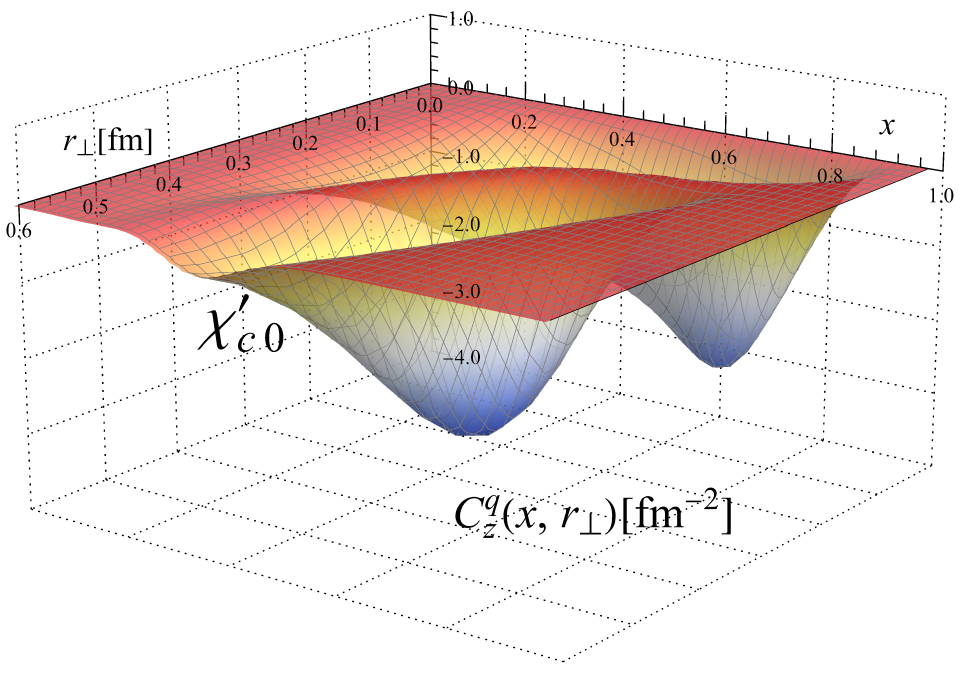}}
    \subfigure[\label{fig:SOCPDF_Bc_C_1S_N32_coordinates}]{\includegraphics[width=0.35\textwidth]{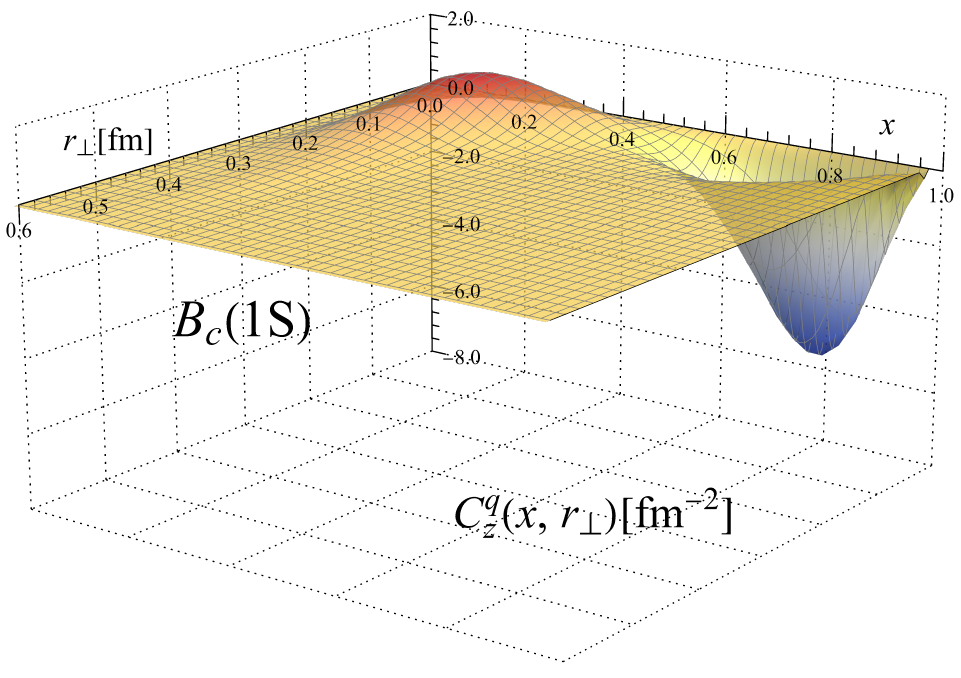}}
    \subfigure[\label{fig:SOCPDF_Bc_C_2S_N32_coordinates}]{\includegraphics[width=0.35\textwidth]{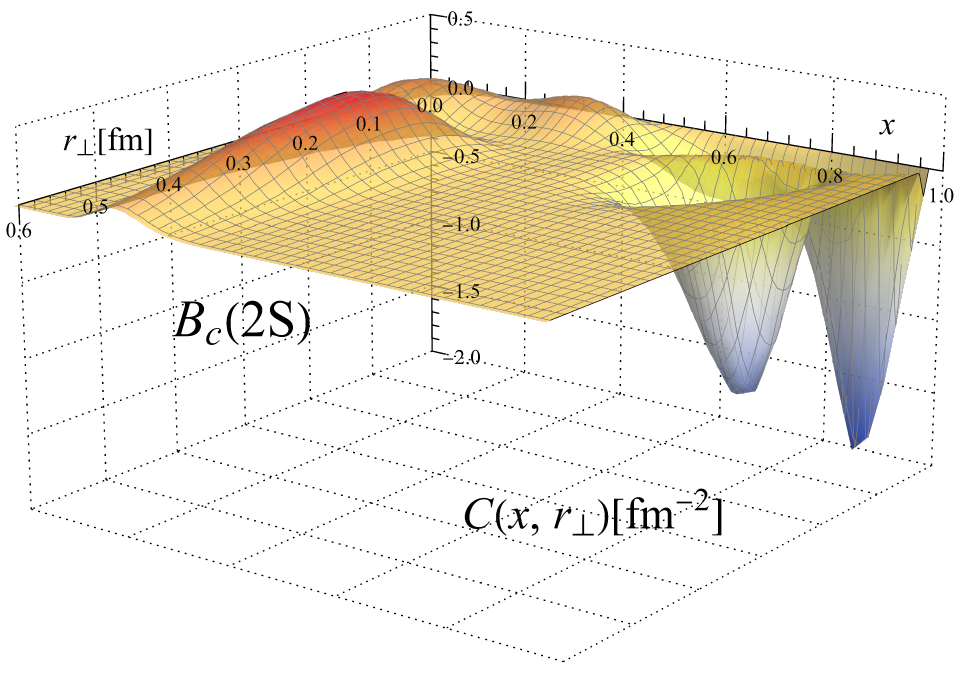}}
    \subfigure[\label{fig:SOCPDF_Bc_C_1P_N32_coordinates}]{\includegraphics[width=0.35\textwidth]{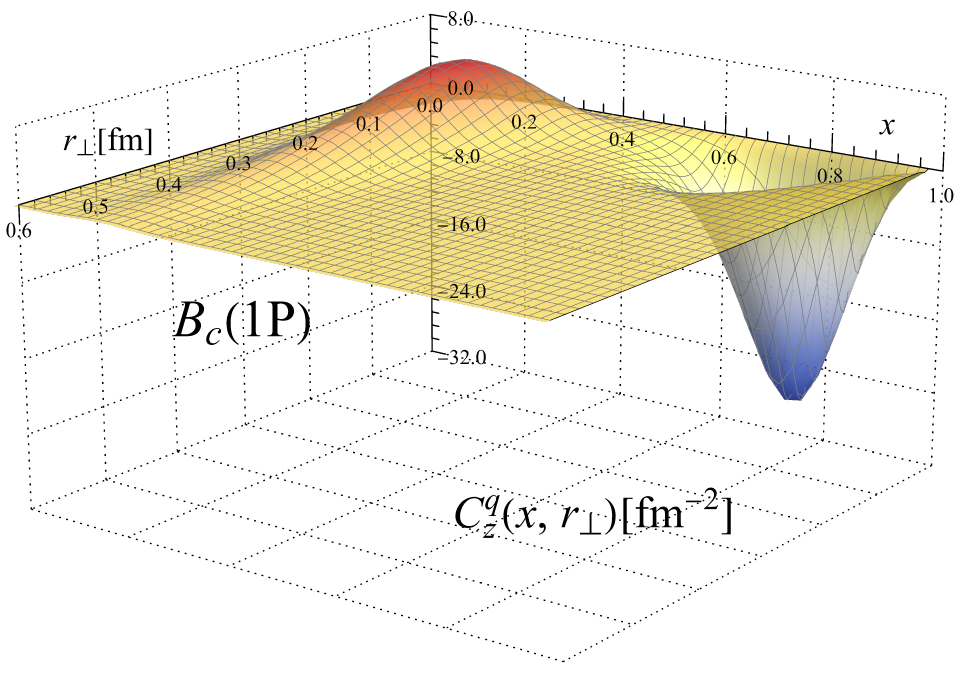}}
    \subfigure[\label{fig:SOCPDF_Bc_C_2P_N32_coordinates}]{\includegraphics[width=0.35\textwidth]{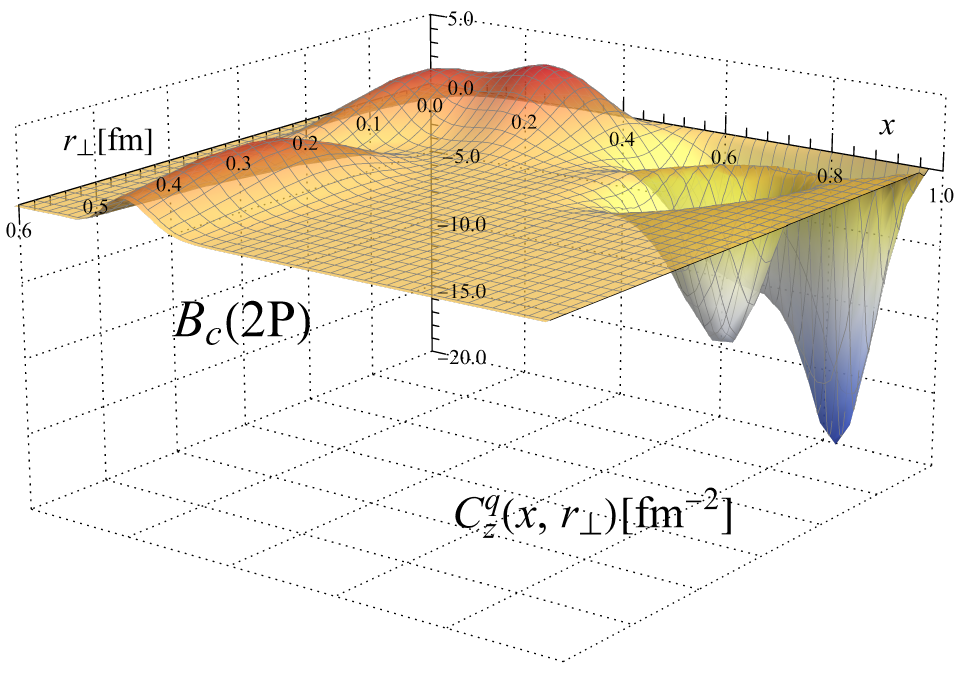}}
    \caption{Three-dimensional parton distributions of the spin-orbit correlation, $C^q_z(x, r_\perp)$, for (a) $\eta_c$, (b) $\eta'_c$, (c) $\chi_{c0}$, and (d) $\chi_{c0}'$; as well as (e) $B_c(1S)$, (f) $B_c(2S)$, (g) $B_c(1P)$, and (h) $B_c(2P)$. For the charmonium states in (a--d), only the quark contribution is presented. For the $B_c$ states in (e--h), the total contribution is shown.
    }
    \label{fig:SOCPDF_coordinates}
\end{figure*}

The parton distribution of the SOC along the longitudinal direction, $C_z^q(x)$, is illustrated in Fig.~\ref{fig:SOCPDF}. For the charge-symmetric charmonium system, the distributions for both the quark and antiquark peak around $x=1/2$. Conversely, the unequal constituent masses in the $B_c$ meson cause a relative shift in the peaks of the $b$- and $c$-quark distributions, yielding a distinctive zig-zag profile. To provide a complete spatial picture, the fully unintegrated 3D distribution of the SOC on the light front is visualized in Fig.~\ref{fig:SOCPDF_coordinates}.

\begin{figure*}[!t]
    \centering
    \subfigure[\label{fig:SOC_etac_F_1S_N816}]{\includegraphics[width=0.4\textwidth]{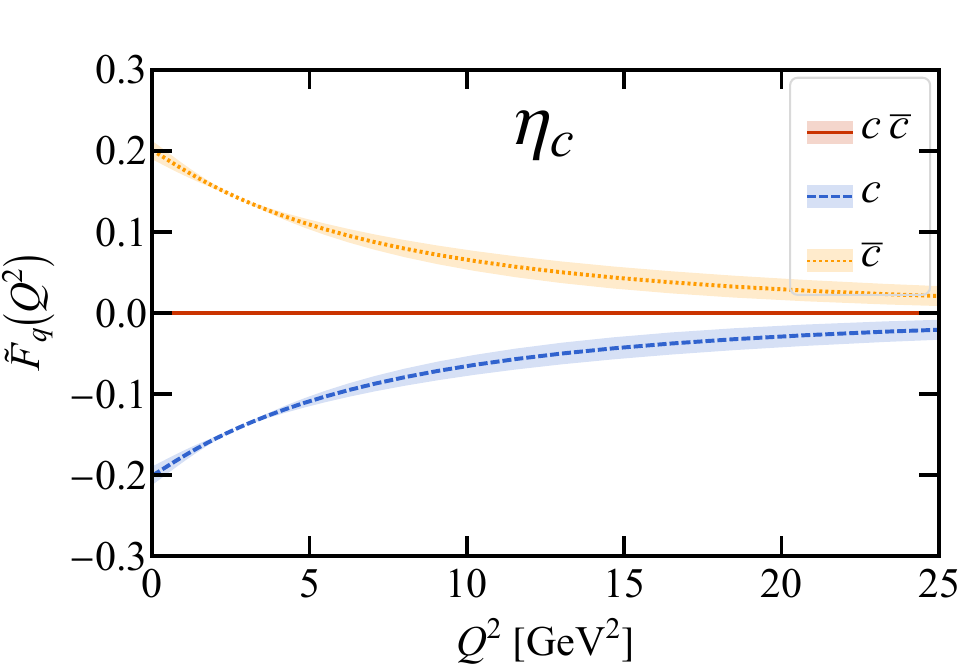}}
    \subfigure[\label{fig:SOC_Bc_F_1S_N2432}]{\includegraphics[width=0.41\textwidth]{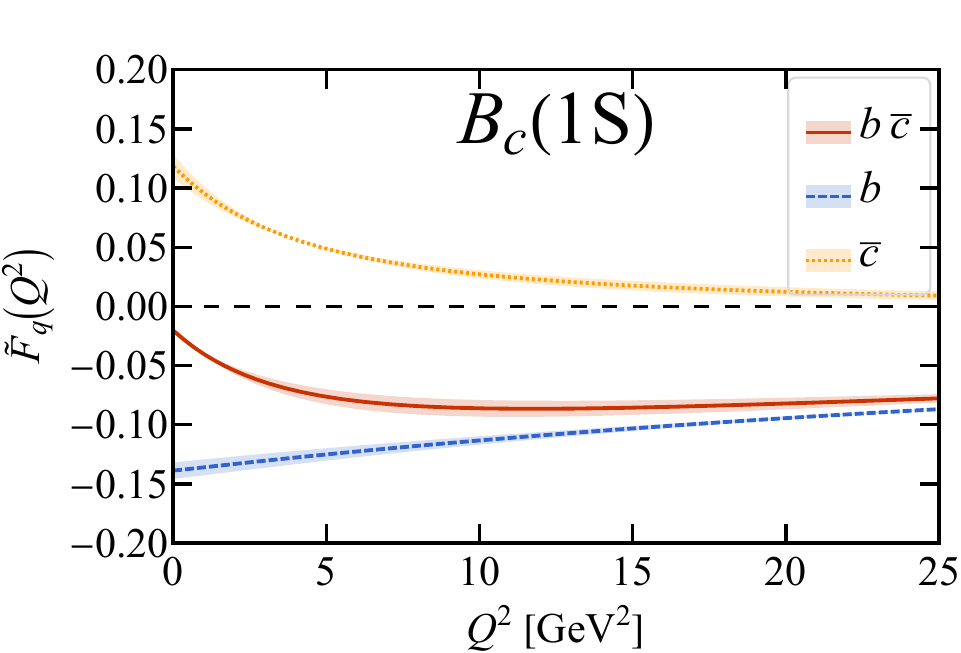}}\\
    \subfigure[\label{fig:SOC_chic0_F_1P_N816}]{\includegraphics[width=0.4\textwidth]{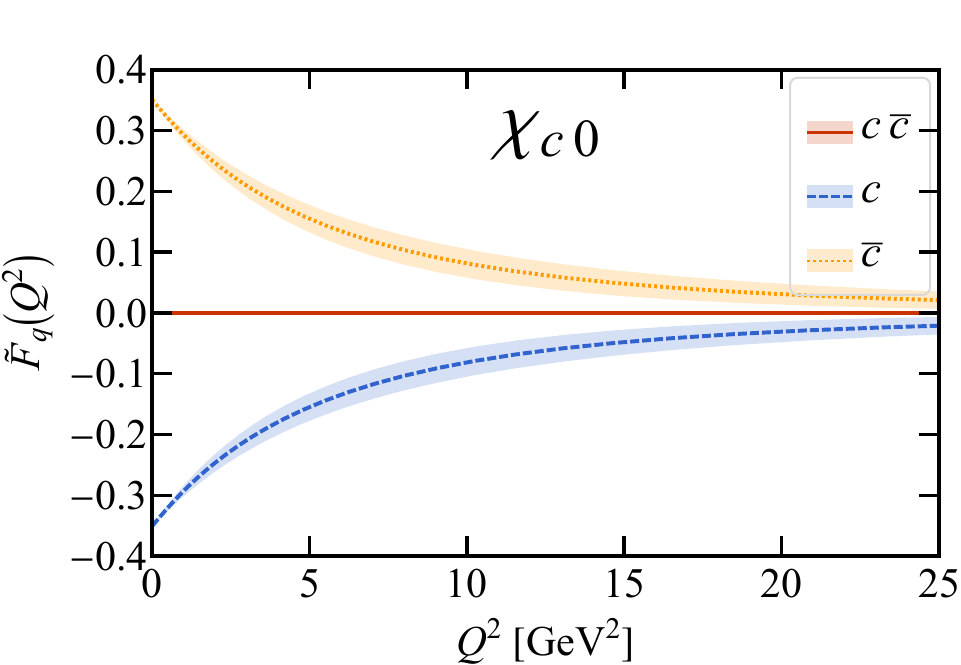}}
    \subfigure[\label{fig:SOC_Bc_F_1P_N2432}]{\includegraphics[width=0.405\textwidth]{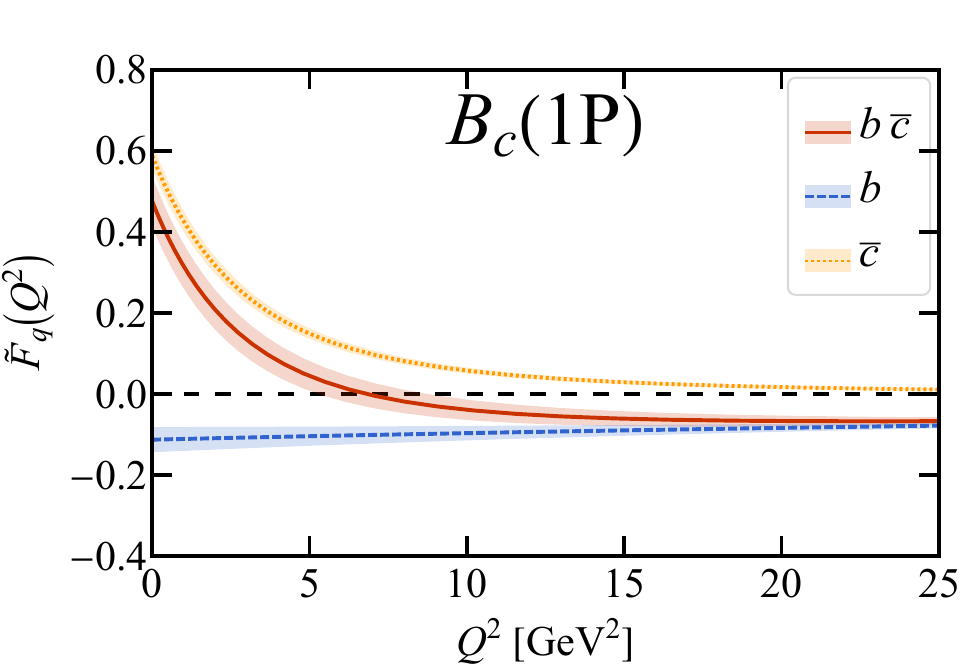}}
    \caption{Calculated FF $\widetilde F_q(Q^2)$ across selected S-wave and P-wave systems. Panels display the results for (a) the S-wave charmonium $\eta_c$, (b) the S-wave $B_c$ meson, (c) the P-wave charmonium $\chi_{c0}$, and (d) the P-wave $B_c$ meson. The accompanying shaded bands reflect the theoretical sensitivity to the basis truncation (see text).}
    \label{fig:SOC_eatc_Bc_F_1S}
\end{figure*}

\begin{figure*}
    \centering
    \subfigure[\label{fig:SOC_etac_trueFofq_1S2S3S_N816}]{\includegraphics[width=0.4\textwidth]{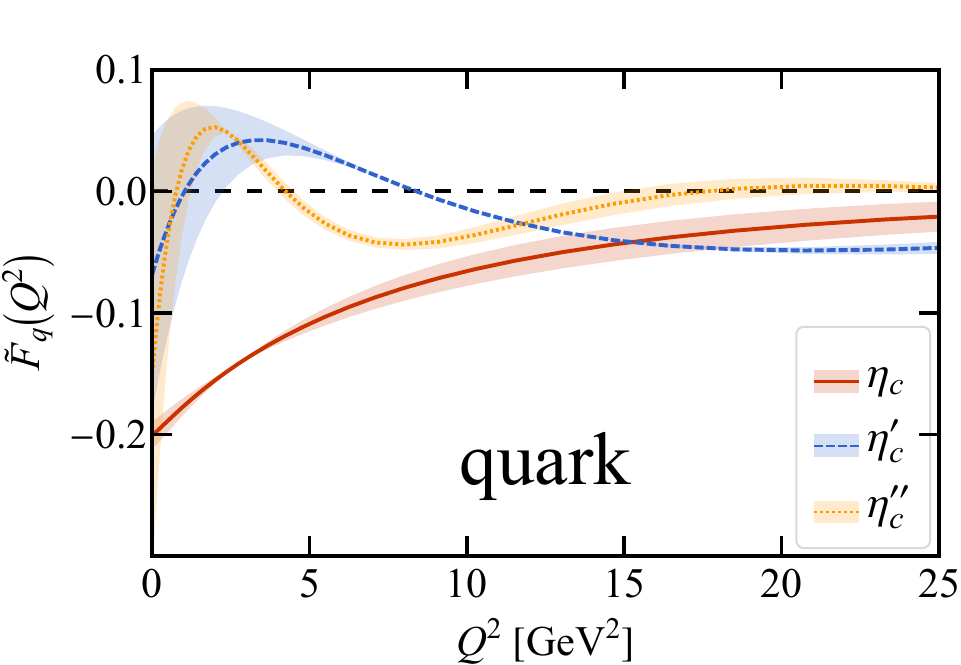}}
    \subfigure[\label{fig:SOC_chic0_trueFofq_1P2P_N816}]{\includegraphics[width=0.4\textwidth]{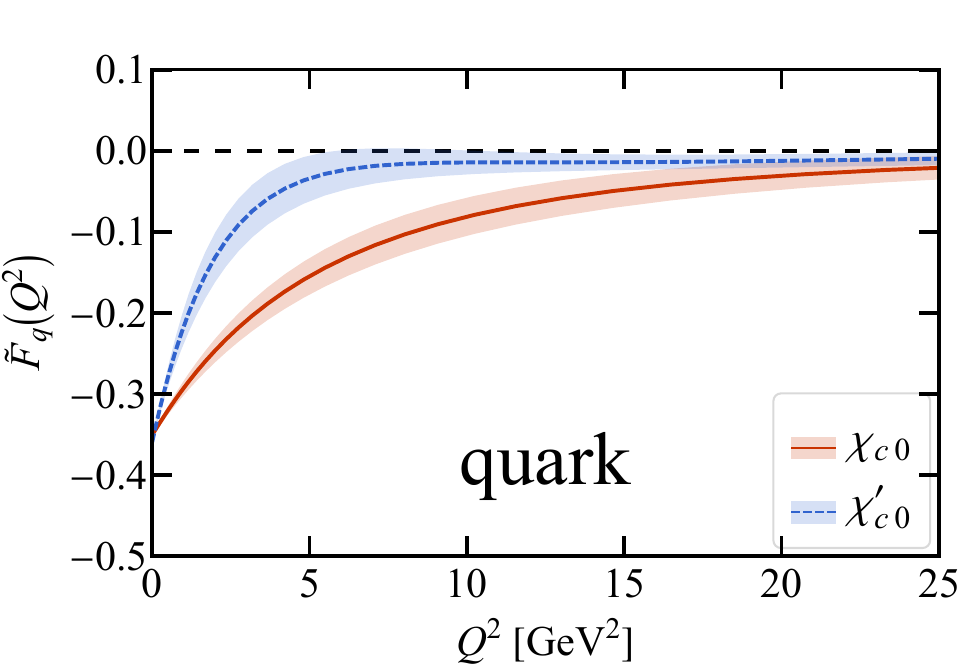}}
    \subfigure[\label{fig:SOC_Bc_trueF_1S2S_N2432}]{\includegraphics[width=0.4\textwidth]{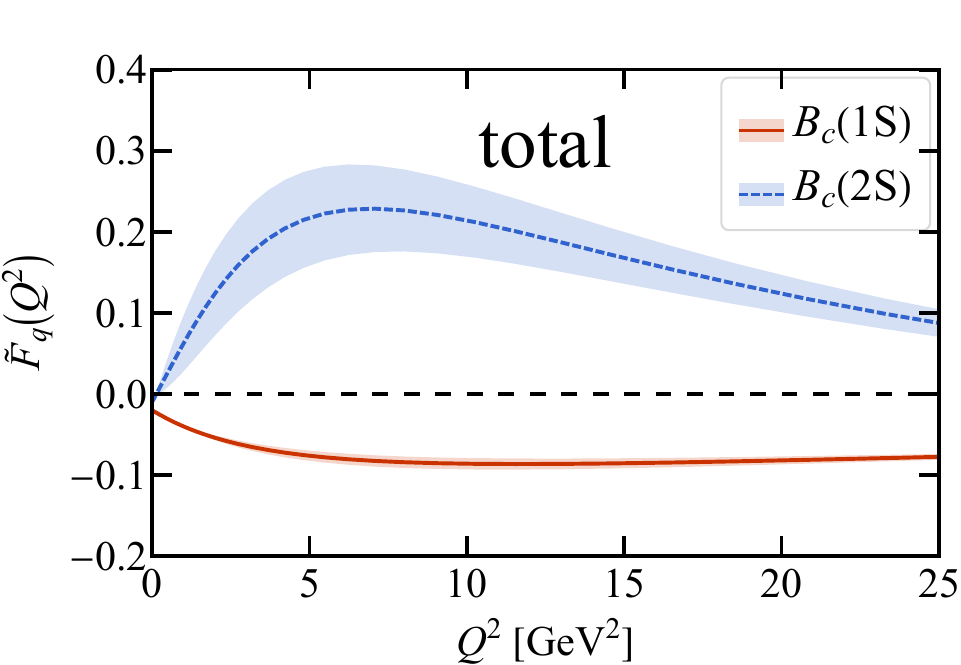}}
    \subfigure[\label{fig:SOC_Bc_trueF_1P2P_N2432}]{\includegraphics[width=0.4\textwidth]{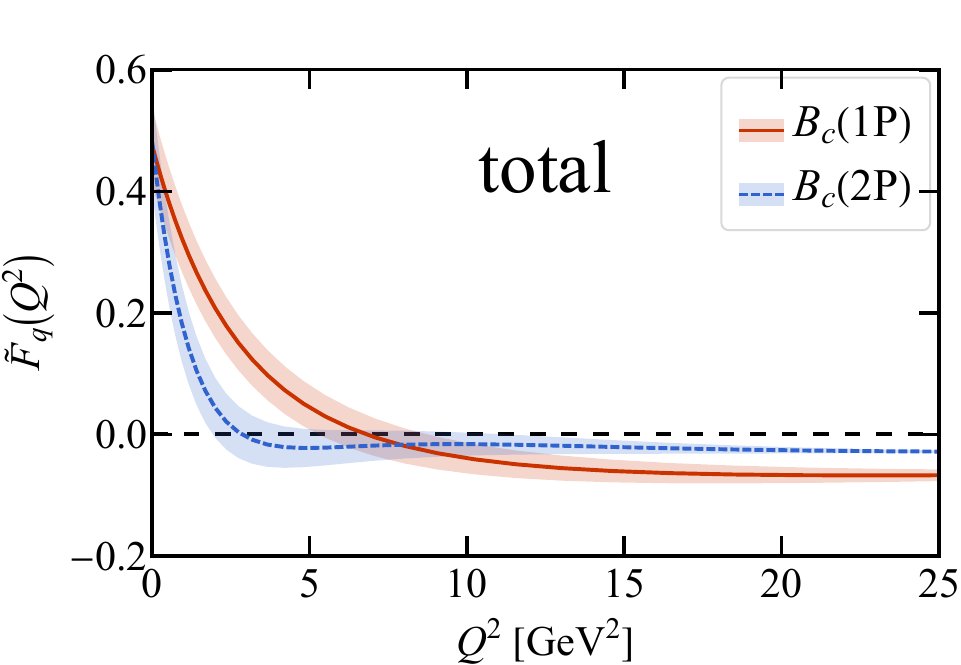}}
    \caption{FF $\widetilde F_q(Q^2)$ evaluated across selected radial excitations. The panels are organized by system and orbital angular momentum: (a) S-wave charmonia ($\eta_c$, $\eta'_c$, and $\eta''_c$), (b) P-wave charmonia ($\chi_{c0}$ and $\chi_{c0}'$), (c) S-wave $B_c$ mesons ($B_c(1S)$ and $B_c(2S)$), and (d) P-wave $B_c$ mesons ($B_c(1P)$ and $B_c(2P)$). The accompanying shaded bands indicate the theoretical uncertainty associated with the basis truncation sensitivity (see text).}
    \label{fig:SOC_Fofq_N816}
\end{figure*}

The results for the FF $\widetilde F_q(Q^2)$ are shown in Figs.~\ref{fig:SOC_eatc_Bc_F_1S}--\ref{fig:SOC_Fofq_N816}. The FF $\widetilde F_q(Q^2)$ for the 1S charmonium state $\eta_c$ is presented in Fig.~\ref{fig:SOC_etac_F_1S_N816}, while the corresponding results for the $B_c(1\text{S})$ meson are shown in Fig.~\ref{fig:SOC_Bc_F_1S_N2432}. Figures~\ref{fig:SOC_chic0_F_1P_N816} and \ref{fig:SOC_Bc_F_1P_N2432} display the results for the P-wave states, $\chi_{c0}$ and $B_c(1\text{P})$, which exhibit significantly less sensitivity to the basis truncation. 
In the covariant limit, one expects $\widetilde F_q(0) = C_z^q$. Although the P-wave states approximately satisfy this relation, the resulting $\widetilde F_q(0)$ for the S-wave states is an order of magnitude larger than $C_z^q$. This discrepancy indicates a substantial spurious contribution within the S-waves.
Furthermore, Fig.~\ref{fig:SOC_Fofq_N816} compares the quark FFs $\widetilde F_q(Q^2)$ across different radial excitations. Specifically, Fig.~\ref{fig:SOC_etac_trueFofq_1S2S3S_N816} illustrates $\widetilde F_q(Q^2)$ for the S-wave charmonia ($\eta_c$, $\eta_c'$, and $\eta_c''$). 
Figure~\ref{fig:SOC_chic0_trueFofq_1P2P_N816} depicts the corresponding quark FFs for the P-wave charmonia ($\chi_{c0}$ and $\chi_{c0}'$). Finally, the total FF $\widetilde F_q(Q^2)$ for the various radially excited $B_c$ states is summarized in Figs.~\ref{fig:SOC_Bc_trueF_1S2S_N2432}--\ref{fig:SOC_Bc_trueF_1P2P_N2432}.

\section{Summary and outlooks}\label{sec:conclusion}

In this work, we have systematically investigated the spin-orbit correlation (SOC) of quarks using a relativistic quantum many-body approach. By deriving the exact many-body representation of $C_z^q$, we demonstrated that it evaluates to twice the longitudinal SOC on the light front, i.e., $\hat C_z^q = 2 \sum_i L_i^z S_i^z$. This physical interpretation extends naturally to the spatial domain, where the density $\mathcal C_z^q(r_\perp)$ is identified as twice the SOC one-body density. Finally, we provided a rigorous partonic interpretation of the SOC by linking it directly to the formalism of generalized parton distributions (GPDs).

Prior work by Lorcé et al. established a fundamental link between the SOC $C_z^q$ and the hadronic matrix element of the P-odd energy-momentum tensor (EMT) $\widetilde T^{\mu\nu}_q$. Nevertheless, tensions persist in the literature concerning the correct procedure to extract the associated FF $\widetilde F_q$ in practical calculations. In this work, we resolved this ambiguity by conducting a covariant light-front dynamics (CLFD) analysis, yielding the complete parametrization of the matrix element $\langle p' | \widetilde T^{\mu\nu}_q(0)| p\rangle$ for hadronic state described by partonic degrees of freedom. We rigorously demonstrated that the genuine physical FF $\widetilde F_q$ is accessed exclusively through the antisymmetric P-odd EMT component, $\widetilde T^{[+i]}_q$. Our framework not only generalizes the decomposition introduced by Tan and Lyu but also smoothly recovers the standard covariant decomposition in the appropriate limit. Consequently, this extraction method provides a robust theoretical tool for practical light-front calculations where manifest covariance is broken.

As a practical application of this formalism, we computed the SOC for charmonia and $B_c$ mesons. As anticipated, the numerical results are consistent with estimates based on the non-relativistic quark model (NRQM). We also compared the transverse and longitudinal SOC distributions in detail across various S-wave, P-wave, and D-wave quarkonia, including radially excited states, which exhibit distinct and intriguing spatial topologies. 

Experimental access to the SOC presents a unique phenomenological challenge. Formally, the most direct pathway to the SOC is through the framework of GPDs. As we have demonstrated, the physical P-odd EMT FF rigorously connects to specific twist-3 GPDs, providing a profound 3D tomographic map of spin-orbit entanglement in impact-parameter space. For stable hadrons like the nucleon, such GPDs are systematically probed via deeply virtual Compton scattering (DVCS). However, the unstable nature of heavy quarkonia precludes their use as fixed targets, rendering standard DVCS measurements unfeasible. Nevertheless, this exact mapping to GPDs remains highly advantageous: it tightly links the macroscopic SOC to the underlying microscopic light-front wave functions (LFWFs). Consequently, precision measurements of hard exclusive processes, such as two-photon fusion ($\gamma^* \gamma^* \to \eta_c, \chi_{c0}$) or exclusive double charmonium production at $e^+ e^-$ colliders (e.g., BES III, Belle II, and the proposed Super Tau-Charm Factory), can strictly constrain these phenomenological LFWFs, providing a robust, data-driven method to reconstruct the quarkonium GPDs and extract the internal SOC. 

Complementary to this spatial GPD picture, the partonic SOC can be probed comprehensively in full phase space through Generalized Transverse Momentum Distributions (GTMDs). As the exact quantum mechanical phase-space distributions of the bound state, GTMDs house the most direct macroscopic representation of the SOC. Recently, as proposed by Bhattacharya et al., highly promising avenues to directly access the proton's SOC via GTMDs at the EIC have been established using exclusive dijet production ($\gamma^* p \to p' + \text{jet} + \text{jet}$) and semi-inclusive diffractive deep inelastic scattering (SIDDIS) \cite{Bhattacharya:2024sck}.

Our current numerical evaluation is restricted to the valence quark sector, which serves as a robust approximation for heavy quarkonia. However, the SOC observable is fundamentally capable of probing the full many-body content of QCD bound states. Consequently, investigating the SOC contributions from sea quarks and gluons remains a primary goal for future study. For scalar mesons, a particularly intriguing paradox arises regarding these gluon contributions. On one hand, the many-body representation can be directly extended to incorporate gluons, viz. $\hat C_z^{q,g} = 2\sum_{i\in q, g} L_i^z S_i^z$. On the other hand, Lorcé and Song demonstrated the absence of an appropriate gluonic operator for the P-odd EMT, implying that the macroscopic gluonic contribution to scalar mesons must strictly vanish. Reconciling this apparent tension between the microscopic many-body picture and the macroscopic EMT formalism will be a critical step toward unraveling the complex gluonic structure of QCD.

\section*{Acknowledgements}

%We wish to thank xxx for fruitful discussions. 
%
 This work is supported by the National Natural Science Foundation of China (NSFC) under Grant No.~12375081, and from the Chinese Academy of Sciences under Grant No.~YSBR-101.  A portion of the computational resources were also provided by Dongjiang Yuan Intelligent Computing Center. 

\appendix

\section{Many-body kinematics on the light front}\label{sec:light-front_kinematics}

For an $n$-body light-front quantized Fock state 
$$
|\{p_i^+, \vec p_{i\perp}, s_i\}\rangle \equiv |p_1^+, \vec p_{1\perp}, s_1, p_2^+, \vec p_{2\perp}, s_2, \cdots, p_n^+, \vec p_{n\perp}, s_n\rangle,$$ we introduce the relative momenta as,
\begin{equation}
x_i = p_i^+/P^+, \quad \vec k_{i\perp} = \vec p_{i\perp} - x_i \vec P_\perp,
\end{equation}
where, $P^+ = \sum_i p_i^+$, and $\vec P_\perp = \sum_i \vec p_{i\perp}$. The relative momenta satisfy
\begin{equation}
\sum_i x_i = 1, \quad \sum_i \vec k_{i\perp} = 0.
\end{equation}
The light-front wave functions only depend on these relative variables, viz. 
\begin{equation}
\langle \{p_i^+, \vec p_{i\perp}, s_i\} | \psi \rangle = 
\psi(\{x_i, \vec k_{i\perp}, s_i\}). 
\end{equation}
The $n$-body measure is defined as, 
\begin{multline}
\int\big[ \dd x_i \dd^2k_{i\perp} \big]_n = \frac{1}{S_n}\prod_{i=1}^{n}
\int \frac{\dd x_i}{2x_i}\frac{\dd^2k_{i\perp}}{(2\pi)^3} \\
\times 2\delta(\sum_i x_i - 1) (2\pi)^3\delta^2(\sum_i \vec k_{i\perp} ) 
\end{multline}
where, $S_n$ is the symmetry factor. 

The transverse coordinate-space wave function is defined as,
\begin{multline}\label{eqn:intrinsic_coordinate_lfwf}
\widetilde\psi_n(\{x_i, \vec r_{i\perp}, s_i\}) = \int \big[\dd^2k_{i\perp}\big]_n e^{-i\sum_i\vec k_{i\perp}\cdot \vec r_{i\perp}} \\
\times \psi_n(\{x_i, \vec k_{i\perp}, s_i\}). 
\end{multline}
Here, we have denoted, 
\begin{equation}
\int\big[\dd^2 k_{i\perp}\big]_n = \prod_{i=1}^{n} \int \frac{\dd^2k_{i\perp}}{(2\pi)^2} (2\pi)^2\delta^2(\sum_i \vec k_{i\perp}).
\end{equation}
Finally, the $n$-body measure of the transverse coordinates is defined as, 
\begin{multline}
\int\big[\dd x_i \dd^2 r_{i\perp} \big]_n 
= \frac{1}{S_n} \prod_{i=1}^n \int \frac{\dd x_i}{4\pi x_i} \dd^2r_{i\perp} \\
\times 4\pi \delta(\sum_i x_i - 1) \delta^2(\sum_i x_i \vec r_{i\perp}).
\end{multline}

\section{Hadronic matrix elements}\label{sec:hadron_matrix_elements}

The covariant decomposition of the parity-odd tensor matrix element for spin-0 hadrons is given by Eqs.~(\ref{eq:P-odd_Tmunu_asym}--\ref{eq:P-odd_Tmunu_sym}): 
\begin{align}
\langle p' | \widetilde T^{[\mu\nu]}_q(0) |p \rangle = \,&  -i \varepsilon^{\mu\nu \Delta P} \widetilde F_q(t) 
 + \varepsilon^{\mu\nu PN}  \widetilde A_1(t) \nonumber \\
 & + i\varepsilon^{\mu\nu\Delta N} \widetilde A_2(t) \\
\langle p' | \widetilde T^{\{\mu\nu\}}_q(0) |p \rangle =\,& 
-iP^{\{\mu}\varepsilon^{\nu\} P\Delta N} \widetilde S_1(t) 
 + \Delta^{\{\mu}\varepsilon^{\nu\}P\Delta N} \widetilde S_2(t) \nonumber \\
& + iN^{\{\mu}\varepsilon^{\nu\} P \Delta N} \widetilde S_3(t)
\end{align}
In the symmetric Drell-Yan frame (\(\Delta^+=\Delta^-=0\), \(\vec{P}_\perp=0\)), the light-front components are: 
\begin{widetext}
\begin{align}
& \langle p' | \widetilde T^{[+-]}_q | p\rangle  =  \langle p' | \widetilde T^{[++]}_q | p\rangle  = \langle p' | \widetilde T^{[--]}_q | p\rangle  = \langle p' | \widetilde T^{[11]}_q | p\rangle  = \langle p' | \widetilde T^{[22]}_q | p\rangle  = 0, \\
& \langle p' | \widetilde T^{[+i]}_q | p\rangle = -i P^+ \epsilon^{ij}_T \Delta^j_\perp \widetilde F (\Delta_\perp^2), \\
& \langle p' |\widetilde T^{[12]}_q | p\rangle = M^2 A_1(\Delta_\perp^2), \\ 
& \langle p' |\widetilde T^{[-i]}_q | p\rangle = -\frac{i \epsilon^{ij}_T \Delta^j_\perp}{P^+}\Big\{(M^2+\frac{1}{4}\Delta^2_\perp) \widetilde F (\Delta_\perp^2)  + 2M^2 A_2(\Delta_\perp^2)  \Big\}, \\
& \langle p' | \widetilde T^{\{+-\}}_q | p\rangle  = \langle p' | \widetilde T^{\{++\}}_q | p\rangle = \langle p' | \widetilde T^{\{--\}}_q | p\rangle = 0, \\
& \langle p' | \widetilde T^{\{+i\}}_q | p\rangle = -i M^2 P^+ \epsilon^{ij}_T \Delta^j_\perp \widetilde S_1 (\Delta_\perp^2), \\
& \langle p' | \widetilde T^{\{-i\}}_q | p\rangle = M^2 \frac{i \epsilon^{ij}_T \Delta^j_\perp}{P^+}\Big\{-(M^2+\frac{1}{4}\Delta^2_\perp) \widetilde S_1 (\Delta_\perp^2)  + 2M^2 S_3(\Delta_\perp^2)  \Big\}, \\
& \langle p' | \widetilde T^{\{11\}}_q | p\rangle = 2\Delta_\perp^1 \Delta_\perp^2 M^2S_2(\Delta^2_\perp), \\
& \langle p' | \widetilde T^{\{22\}}_q | p\rangle = -2\Delta_\perp^1 \Delta_\perp^2 M^2S_2(\Delta^2_\perp), \\
& \langle p' | \widetilde T^{\{12\}}_q | p\rangle = \big[(\Delta_\perp^2)^2 - (\Delta_\perp^1)^2\big]M^2S_2(\Delta^2_\perp).
\end{align}
\end{widetext}
Here, the transverse Levi-Civita symbol is defined by $\varepsilon^{12}_T = +1$.

\end{document}